\documentclass[a4paper,fleqn,usenatbib]{mnras}

\usepackage{newtxtext,newtxmath}
\usepackage[T1]{fontenc}
\usepackage{ae,aecompl}
\usepackage{pdflscape}
\usepackage{threeparttable}

\usepackage{graphicx}	% Including figure files
\usepackage{amsmath}	% Advanced maths commands
\usepackage{amssymb}	% Extra maths symbols

\title[ALMA observations of a proto-brown dwarf]{ALMA reveals a pseudo-disc in a proto-brown dwarf}

\author[Riaz et al.]{B. Riaz,$^{1}$\thanks{E-mail: briaz@usm.lmu.de}
M. N. Machida,$^{2}$
D. Stamatellos$^{3}$
\\
$^{1}$  Universit\"ats-Sternwarte M\"unchen, Ludwig-Maximilians-Universit\"at, Scheinerstr.~1, 81679 M\"unchen, Germany \\
$^{2}$ Department of Earth and Planetary Sciences, Faculty of Sciences, Kyushu University, Fukuoka, Japan \\
$^{3}$ Jeremiah Horrocks Institute for Mathematics, Physics \& Astronomy, University of Central Lancashire, Preston, PR1 2HE, UK  \\
}

\date{Accepted XXX. Received YYY; in original form ZZZ}

\pubyear{2019}

\begin{document}
\label{firstpage}
\pagerange{\pageref{firstpage}--\pageref{lastpage}}
\maketitle

\begin{abstract}

We present the observational evidence of a pseudo-disc around the proto-brown dwarf Mayrit 1701117, the driving source of the large-scale HH~1165 jet. Our analysis is based on ALMA $^{12}$CO (2-1) line and 1.37 mm continuum observations at an angular resolution of $\sim$0.4$\arcsec$. The pseudo-disc is a bright feature in the CO position-velocity diagram (PVD), elongated in a direction perpendicular to the jet axis, with a total (gas+dust) mass of $\sim$0.02 M$_{\sun}$, size of 165-192 AU, and a velocity spread of $\pm$2 km s$^{-1}$. The large velocity gradient is a combination of infalling and rotational motions, indicating a contribution from a pseudo-disc and an unresolved inner Keplerian disc. There is weak emission detected in the H$_{2}$CO (3-2) and N$_{2}$D$^{+}$ (3-2) lines. H$_{2}$CO emission likely probes the inner Keplerian disc where CO is expected to be frozen, while N$_{2}$D$^{+}$ possibly originates from an enhanced clump at the outer edge of the pseudo-disc. We have considered various models (core collapse, disc fragmentation, circum-binary disc) that can fit both the observed CO spectrum and the position-velocity offsets. The observed morphology, velocity structure, and the physical dimensions of the pseudo-disc are consistent with the predictions from the core collapse simulations for brown dwarf formation. From the best model fit, we can constrain the age of the proto-brown dwarf system to be $\sim$30,000-40,000 yr. A comparison of the H$_{2}$ column density derived from the CO line and 1.37 mm continuum emission indicates that only about 2\% of the CO is depleted from the gas phase.

%The source size for M1701117 as measured from the 1.37 mm continuum image is $\sim$270 AU. 
%The CO line PVD along the jet axis shows a compact ($<$150 AU) unresolved molecular outflow. 

%The very small value of the dust opacity index $\beta$$\sim$0.1 measured for M1701117 can be explained by the effect of dust grain growth.

%This object is in transition from Class 0 to Class I, consistent with the finding of a pseudo-disk structure, intermediate between an infalling envelope and a Keplerian disk. 

%the observed CO spectrum shows both infall and Keplerian rotation signatures as well as broad, extended wings indicative of an unresolved molecular outflow. 

%the observed continuum and line emission likely arises from a pseudo-disk component, intermediate between an infalling envelope and a Keplerian disk. 

%We have derived a CO abundance (relative to H$_{2}$) of 8.6$\times$10$^{-9}$ from line radiative transfer modelling of the observed CO spectrum. A comparison of the H$_{2}$ column density derived from the CO gas emission with that derived from 1.37 mm dust continuum emission indicates that $\sim$15\% of the CO is depleted from the gas phase. 

\end{abstract}


%%%%%%%%%%%%%%%%%%%%%%%%%%%%%%%%%%%%%%%%%%%%%

%%%%%%%%%%%%%%%%% BODY OF PAPER %%%%%%%%%%%%%%%%%%

\section{Introduction}
\label{intro}

Low-mass stars and brown dwarfs may form similarly to their Solar-type counterparts from the collapse of a low-mass molecular cloud core (core collapse) or by disc fragmentation, due to gravitational instabilities, followed by ejection (e.g., Whitworth et al. 2007). Core collapse models of low-mass star formation predict the formation of a flattened ``pseudo-disc'' with spatial extents of a few thousand AU and a symmetry axis parallel to the core magnetic field (e.g., Tomisaka 1998; Mellon \& Li 2009; Machida et al. 2011). The pseudo-disc can be described as an infalling envelope with a flattened, disc-like structure, caused by the magnetic field or Lorentz force that changes the geometry of the collapsing cloud. Although the term `pseudo' implies that the infalling envelope material is not supported by the rotation against gravity, the pseudo-disc is still rotating and slowly contracts with rotation. The inner Keplerian disc is embedded in the optically thick region of the pseudo-disc, and the gas in the pseudo-disc infalls onto the Keplerian disc with rotation. Previous interferometric molecular line observations have revealed an infalling motion entangled with weak signatures of rotational motion in such flattened pseudo-discs around Class 0/I protostars (e.g., Yen et al. 2011; Davidson et al. 2011; Takakuwa et al. 2012). Brown dwarf formation via gravitational core collapse has been theoretically explored in the magneto-hydro-dynamic simulations by Machida et al. (2009), and predicts the formation of a pseudo-disc with a scaled-down spatial extent that is ten times smaller than predicted in low-mass star formation. However, the existence of pseudo-discs around proto-brown dwarfs has not been observationally verified to date. 

Here, we report the first evidence of a pseudo-disc in the proto-brown dwarf Mayrit 1701117 (M1701117), located in the vicinity of the $\sigma$ Orionis cluster (410$\pm$60 pc; Gaia DR2; Gaia Collaboration 2018). M1701117 is the driving source of the HH~1165 jet, which is the largest scale Herbig-Haro jet identified to date that is driven by a sub-stellar object (Riaz et al. 2017). We have conducted a high-sensitivity continuum and spectral line survey across the full length of the HH~1165 jet using the Atacama Large Millimeter/submillimeter Array (ALMA) interferometer, with an aim to study the infall and disc kinematics of the driving source, and to investigate molecular outflow activity and the chemistry of the various shock emission knots in the jet. To interpret the observations, we have conducted line modelling using simulations of brown dwarf formation via gravitational core collapse, and the alternative formation mechanism via disc fragmentation followed by ejection. If there are two different formation mechanisms of brown dwarfs (core collapse and ejection), the formation time scale is expected to be different between them. In such a case, we may observe differences in the molecular abundances because it is expected that the thermal and chemical evolution differ between proto-brown dwarfs formed from different formation mechanisms.

The ALMA observations are presented in Sect.~\ref{obs}, and the analysis of line and continuum data is described in Sect.~\ref{analysis}. A comparison of the observations with the core collapse and disc fragmentation models for brown dwarf formation is presented in Sect.~\ref{model-comp} and Sect.~\ref{discussion}.

%in addition to the core collapse formation, we have explored the alternative formation mech . . disk fragmentation and circum-binary disk scenarios. 

\section{Observations}
\label{obs}

 \begin{figure*}
  \centering              
     \includegraphics[width=3.3in]{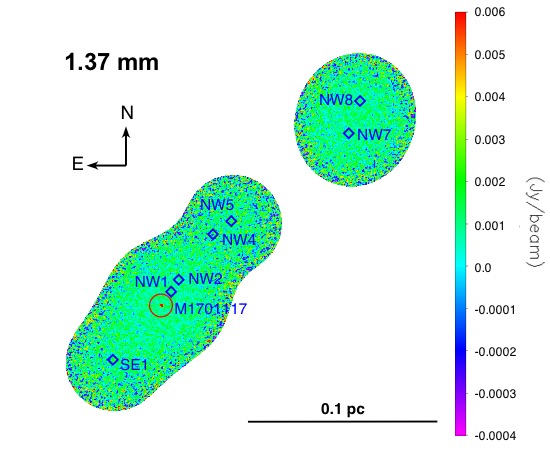}  \hspace{0.1in}
     \includegraphics[width=3.3in]{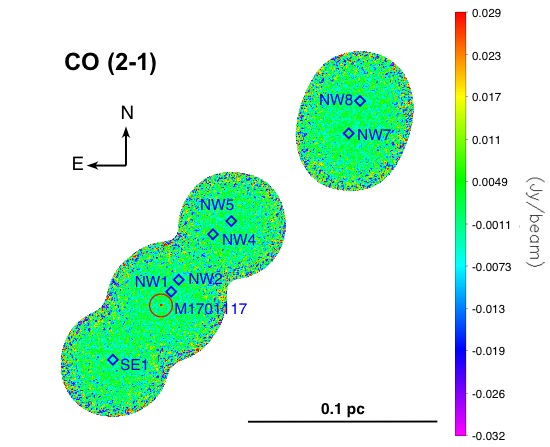}
     \caption{The full map covering the six positions in the 1.37 mm continuum (left) and the $^{12}$CO (2-1) line (right). The positions of the driving source M1701117 and the shock emission knots detected in the optical [SII]$\lambda$6731$\AA$ image are labelled (Riaz et al. 2017). The colour bar shows the flux scale in units of Jy beam$^{-1}$.   }
     \label{maps}
  \end{figure*}

Observations with ALMA were carried out in May, 2016 (PID: 2016.1.01453.S; PI: Riaz). We obtained molecular line and continuum observations in Band 6 (211-275 GHz) at six positions along the HH~1165 jet, one centered on the driving source M1701117, four on the brightest knots in north-west (NW1, NW5, NW7, and NW8), and one on the shocked knot in the south-east SE1 (Fig.~\ref{maps}). The primary beam of the ALMA 12m dishes covers a field of $\sim$22$\arcsec$ in diameter around each of the six pointings, and thus covered additional shocked regions within the field of view. Our chosen configuration was a maximum recoverable scale of $\sim$4$\arcsec$, and an angular resolution of $\sim$0.4$\arcsec$. The spectral setup included the $^{12}$CO, SiO, H$_{2}$CO, SO$_{2}$, H$_{2}$S, HC$_{3}$N, OCS, C$^{18}$O, CH$_{3}$OH, $^{13}$CS, N$_{2}$D$^{+}$ lines. The bandwidth was set to 117.18 MHz, which provides a resolution of 244.14 kHz, or a smoothed resolution of $\sim$0.3 km s$^{-1}$ at 230.5 GHz. In addition to these basebands, one baseband was used for continuum observations where the spectral resolution was set to 976 kHz and the total bandwidth was 1875 MHz. We employed the calibrated data delivered by the EU-ARC. Data analysis was performed on the uniform-weighted synthesized images using the CASA software.

%{\bf this is detection in the para-h2co 3(0,3)-2(0,2) line at 218.2222 GHz. }

There is emission detected in the 1.37 mm continuum and the $^{12}$CO (2-1), para-H$_{2}$CO 3(0,3)-2(0,2), N$_{2}$D$^{+}$ (3-2), and C$^{18}$O (2-1) lines for the driving source M1701117. The brightest detection ($\sim$8-9 $\sigma$) is in the CO line, with a 1-$\sigma$ rms sensitivity of $\sim$3-4 mJy ($\sim$0.3 K), while the remaining lines show a weak detection ($\sim$2-3 $\sigma$), with a 1-$\sigma$ rms sensitivity of $\sim$5-6 mJy ($\sim$0.5 K). The 1-$\sigma$ rms in the 1.37 mm continuum image is 0.8 mJy. No continuum or line emission is detected at a $\geq$2-$\sigma$ level for any of the shocked regions. Figure~\ref{maps} shows the full map covering the six positions in the 1.37 mm continuum and the $^{12}$CO (2-1) line.

%The 1-$\sigma$ rms sensitivity is $\sim$3-4 mJy ($\sim$0.3 K) in the CO line observations, and $\sim$5-6 mJy ($\sim$0.5 K) in the other lines. The 1-$\sigma$ rms in the 1.37 mm continuum image is 0.8 mJy. 

\section{Data Analysis and Results}
\label{analysis}

\subsection{Continuum Map}

 \begin{figure}
  \centering       
       \includegraphics[width=3.3in]{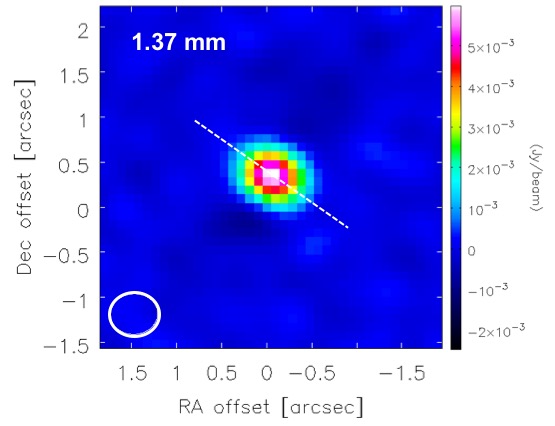}         
     \caption{The 1.37 mm continuum map for M1701117. The white dashed line is along a PA of 55$\degr$. The colour bar shows the flux scale in units of Jy beam$^{-1}$. The beam size is shown at the bottom, left corner. The xy-axes show the position offset relative to the position of M1701117. North is up, east is to the left.  }
     \label{cont}
  \end{figure}

Figure~\ref{cont} shows the 1.37 mm dust continuum image for M1701117. The source is resolved in the continuum and shows a slightly elongated morphology along the north-east to south-west direction. We have used the CASA tasks {\it uvmodelfit} and {\it imfit} to measure the source size, peak position and the continuum flux by fitting one or more elliptical Gaussian components on the image. The peak position of the source is RA (J2000) = 05:40:25.80; Dec (J2000) = -02:48:55.49, with a position uncertainty of $\pm$0.0068$\arcsec$. The source size has a FWHM of 0.53$\arcsec$ along the major axis and 0.45$\arcsec$ along the minor axis, with a position angle (PA) of 55.0$\degr \pm$5.6$\degr$. The uncertainty on the FWHM is 0.02$\arcsec$. The integrated flux is 6.94$\pm$0.41 mJy. The peak flux at the source position is 6.30$\pm$0.22 mJy beam$^{-1}$. Note that the peak flux position is not at (RA, Dec) offset of (0,0) but at (-0.08$\arcsec$, 0.38$\arcsec$) (Fig.~\ref{cont}). The offset is likely due to the $\sim$0.1$\arcsec$ difference in the M1701117 position between the optical and millimeter observations, as well as the beam convolution effect.

%05 40 25.7990148082 -02 48 55.415791769 Optical position
%The source size deconvolved from the beam cannot be accurately determined as the point source is unresolved; an upper limit measured by {\it imfit} is 0.3$\arcsec$ x 0.1$\arcsec$. 
%The source size deconvolved from the beam is about 0.3$\arcsec$ x 0.1$\arcsec$; however, this is likely an upper limit as the point source is unresolved. 
%if IMFIT has been given a previously deconvolved image ? e.g. the output of TCLEAN ? then the integrated flux density it reports is a good measure of the actual integrated flux density of the source, irrespective of the synthesized beam.

Assuming that the dust emission at 1.37 mm is optically thin, the dust continuum mass can be calculated from the millimeter flux density as

\begin{equation}
M_{dust} = \frac{F_{\nu}~d^{2}}{B_{\nu}(T_{dust})~ \kappa_{1.37mm}}~,
\end{equation}

\noindent where $F_{\nu}$ is the flux density at 1.37 mm, $d$ is the distance to the source, $B_{\nu}$ is the Planck function at the dust temperature $T_{dust}$, and $\kappa_{1.37mm}$ is the dust opacity at 1.37 mm. Assuming that the frequency dependence of the dust mass opacity is $\kappa_{\nu}$ = 0.1 $\times$ $\left( \frac{\nu}{10^{12}} \right)^{\beta}$ [cm$^{-2}$ g$^{-1}$] (Beckwith et al. 1990), the dust opacity at 1.37 mm (218.116 GHz) is 0.0218 cm$^{2}$ g$^{-1}$ for $\beta$ = 1.0 or for optically thin emission. The kinetic temperature is expected to be relatively low ($\sim$10 K) throughout the outer and inner envelope layers in proto-brown dwarfs (Machida et al. 2009). Assuming the dust and gas temperature of 10 K, M$_{dust}$ = 0.21$\pm$0.1 M$_{Jup}$. For a gas to dust mass ratio of 100, the total (dust+gas) mass M$_{d+g}$ = 20.98 $\pm$ 1.24 M$_{Jup}$ for M1701117. For a higher $T_{dust}$ = 20 K, the total mass is M$_{d+g}$ = 8 M$_{Jup}$, assuming the same $\beta$ = 1 and gas to dust mass ratio of 100. The uncertainties in independently measuring the gas mass and then derive the gas to dust mass ratio are discussed in Sect.~\ref{discussion}. 

%We note that there are uncertainties in the mass estimates due to the assumptions on the value for the slope $\beta$ of the opacity law, $T_{dust}$, and the gas to dust mass ratio. This is further discussed in Sect.~\ref{discussion}. 

The M$_{d+g}$ derived from ALMA observations with standard parameter sets is a factor of $\sim$2 lower than the (dust+gas) mass of $\sim$36 M$_{Jup}$ derived from the single-dish SCUBA-2 850$\mu$m continuum flux (Riaz et al. 2015). The difference can be explained by the M1701117 source size of $\sim$0.7$\arcsec$ ($\sim$270 AU) as measured from the ALMA 1.37mm continuum image, which is $\sim$20 times smaller than the $\sim$14.5$\arcsec$ beam size of the SCUBA-2 bolocam. The much larger beam size can result in severe contamination from the surrounding medium and enhance the observed emission.

%There are uncertainties in the mass estimates, M$_{d+g}$, made in Sect.~\ref{Continuum Map} due to the assumptions on the value for the slope $\beta$ of the opacity law, $T_{dust}$, and the gas to dust mass ratio. 

%The gas mass derived from the best model fit to the observed CO spectrum is 0.215 M$_{Jup}$ (Sect.~\ref{core}). The dust mass derived independently from the continuum emission is 0.209 M$_{Jup}$. This implies a gas to dust mass ratio of $\sim$1.0. The derived masses would then be lower by a factor of $\sim$100, with M$_{d+g}$ = 0.2 M$_{Jup}$ for $\beta$ = 1.0 and $T_{dust}$ = 10 K. There are, however, degeneracies in the line model fits, and the CO emission is likely optically thick. There is no detection in the optically thin C$^{18}$O isotopologue coincident with the pseudo-disk position (Sect.~\ref{bonus-lines}), which could have been a better tracer of the gas mass than CO. 

%Assuming $\beta \sim$ 1.7 for ISM-like dust, M$_{d+g}$ = 80 M$_{Jup}$

 \begin{figure}
  \centering       
     \includegraphics[width=3.3in]{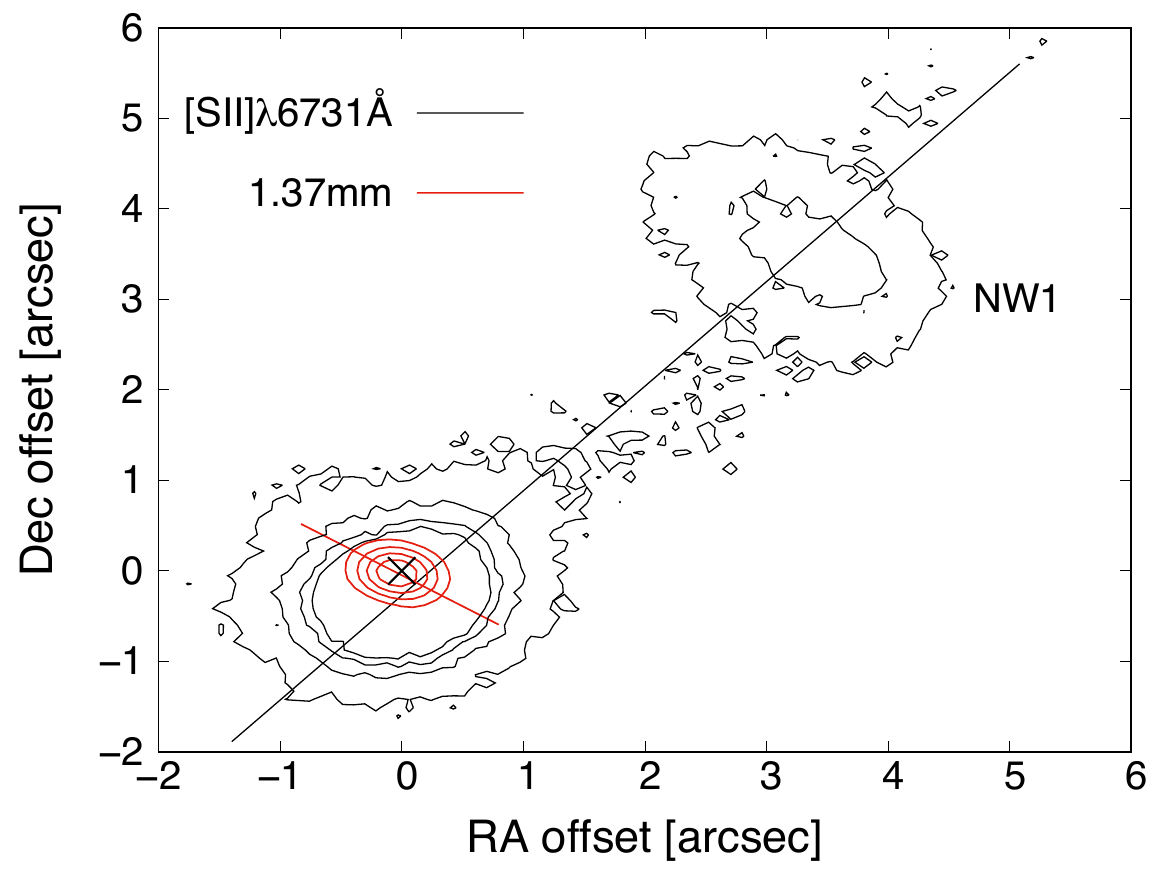}
     \caption{The 1.37 mm continuum map for M1701117 (red contours) overplotted on the optical [SII]$\lambda$6731$\AA$ image of the HH~1165 jet (black contours). Red line is along the {\bf disc axis} (PA = 55$\degr$). Black line is along the {\bf jet axis} (PA  = 320$\degr$). The contours represent the signal-to-noise levels from 2-$\sigma$ to 10-$\sigma$ in steps of 2-$\sigma$. The 1-$\sigma$ rms is 0.8 mJy in the 1.37 mm continuum map and 3$\times$10$^{-16}$ ergs s$^{-1}$ cm$^{-2}$ in the [SII]$\lambda$6731$\AA$ map. The xy-axes show the position offset relative to the 1.37mm continuum position of M1701117 (marked by a black cross). North is up, east is to the left.   }
     \label{SII-cont}
  \end{figure}       
  
 %Cont-mom0 figure: Cont Contours: 0.0012 0.0024 0.00359 0.00479 0.00587

Figure~\ref{SII-cont} shows an overplot of the 1.37 mm continuum map for M1701117 and the optical [SII]$\lambda$6731$\AA$ image of the HH~1165 jet. The [SII] contours are plotted for the location of the driving source and the closest shock emission knot NW1 (Fig.~\ref{maps}). The red solid line in Fig.~\ref{SII-cont} is along the 1.37 mm continuum PA of 55$\degr$. Interestingly, this PA is nearly perpendicular to the HH~1165 jet axis (PA = 320$\degr$$\pm$4$\degr$; black line in Fig.~\ref{SII-cont}). We therefore term the 1.37mm continuum PA (55$\degr$) as the ``disc axis'' and the HH~1165 PA (320$\degr$) as the ``jet axis''. As discussed in Sect.~\ref{momentmaps}, it is also along the disc axis that a velocity gradient is seen in the CO line maps.

\subsection{CO line Channel and Moment Maps}
\label{momentmaps}

\begin{figure*}
  \centering              
     \includegraphics[width=7.5in]{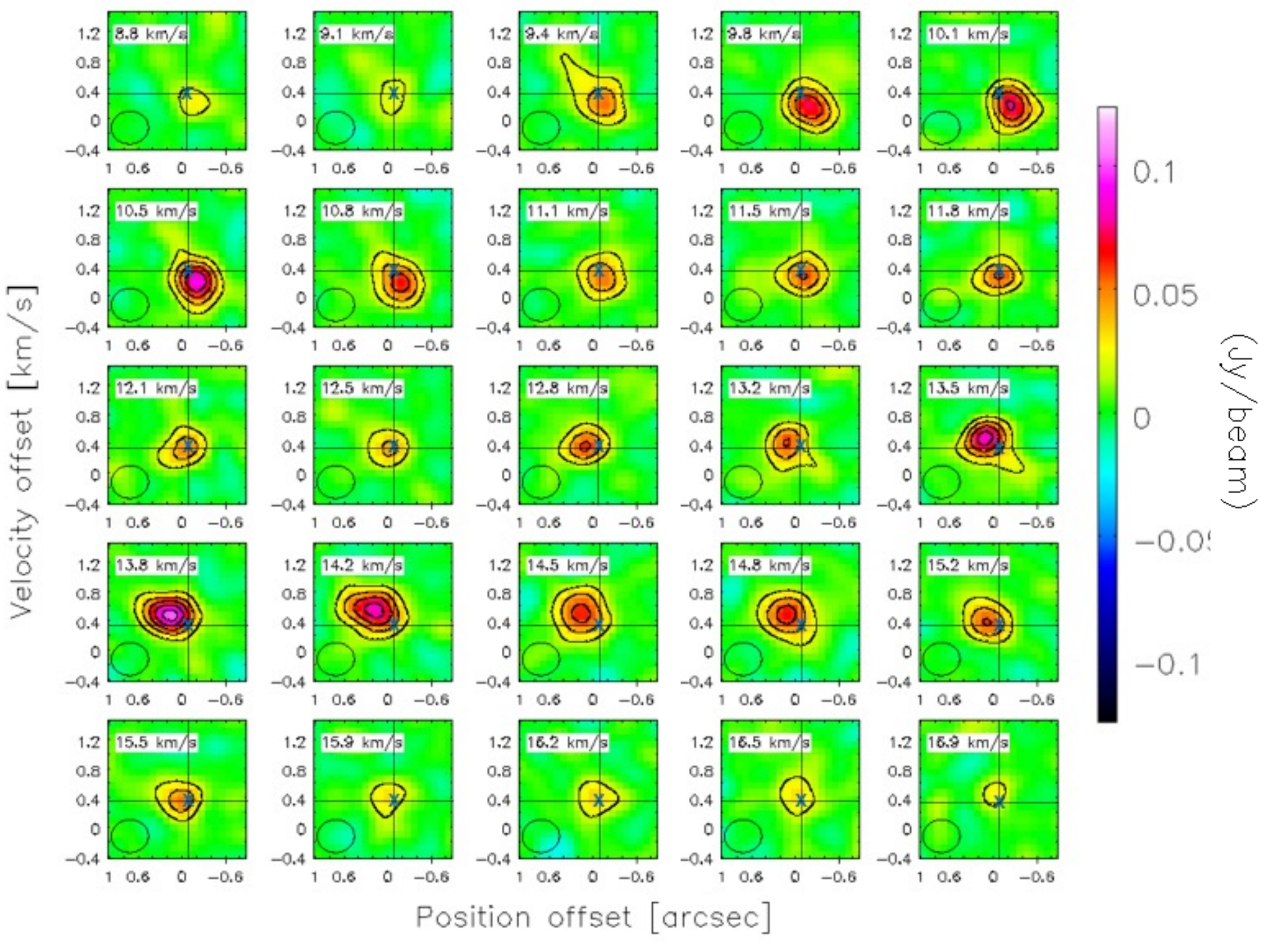}
     \caption{The velocity channel maps in the CO (2-1) line emission for M1701117. The colour bar on the right shows the flux scale in units of Jy beam$^{-1}$. The contour levels are from 2-$\sigma$ to 10-$\sigma$ in steps of 2-$\sigma$. The channel velocity is noted at the top, left corner, and the beam size is shown at the bottom, left corner. The xy-axes show the position offset relative to the position of M1701117, which is marked by a blue cross and dashed black lines. North is up, East is to the left.  }
     \label{ch-maps}
  \end{figure*}

Figure~\ref{ch-maps} shows the CO emission observed in different velocity channel maps. The observed emission in all of the channels has a spatial extent of $\sim$0.4$\arcsec$-0.6$\arcsec$, and it is either marginally resolved or unresolved in the beam size of $\sim$0.4$\arcsec$. The position offset in the channel maps is relative to the position of the source (marked by a cross) as measured from the continuum map. The peak in the CO emission shifts with increasing velocity from the south-west passing through the source position at (0,0) towards the north-east. The emission at or close to the source location is seen at velocities of $\sim$11.8-12.5 km s$^{-1}$. The emission is also compact when it peaks at or close to source location, while it is more extended away from the source at velocities of $\leq$11 km s$^{-1}$ and $\geq$13 km s$^{-1}$. It is noted that the peak emission shows a range in RA offset between approximately -0.5$\arcsec$ to +0.5$\arcsec$, while the offset in Dec is nearly constant at approximately +0.5$\arcsec$. The channel maps thus show a velocity gradient along the north-east to south-west direction at a PA of about 45$\degr$, which is approximately along the disc axis.

\begin{figure*}
  \centering              
     \includegraphics[width=3.3in]{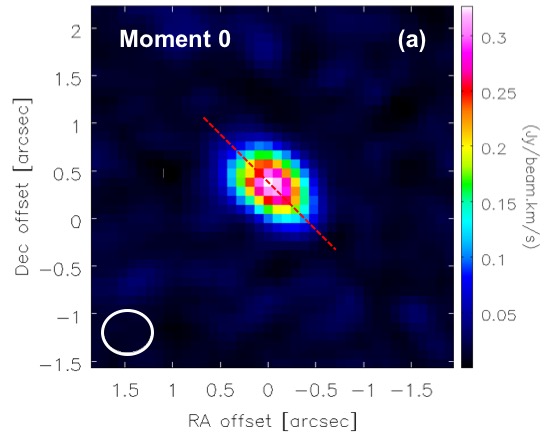}
     \includegraphics[width=3.05in]{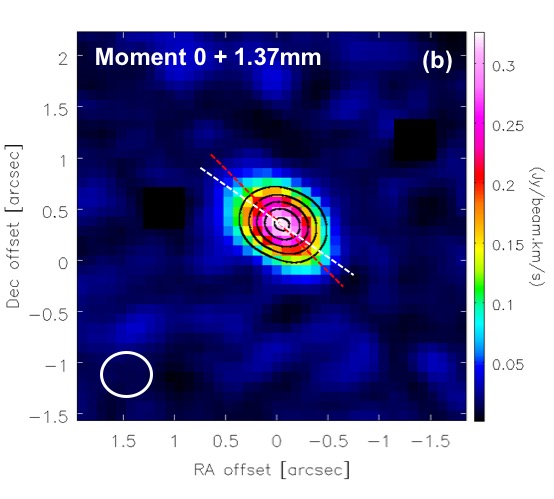}     
     \includegraphics[width=3.1in]{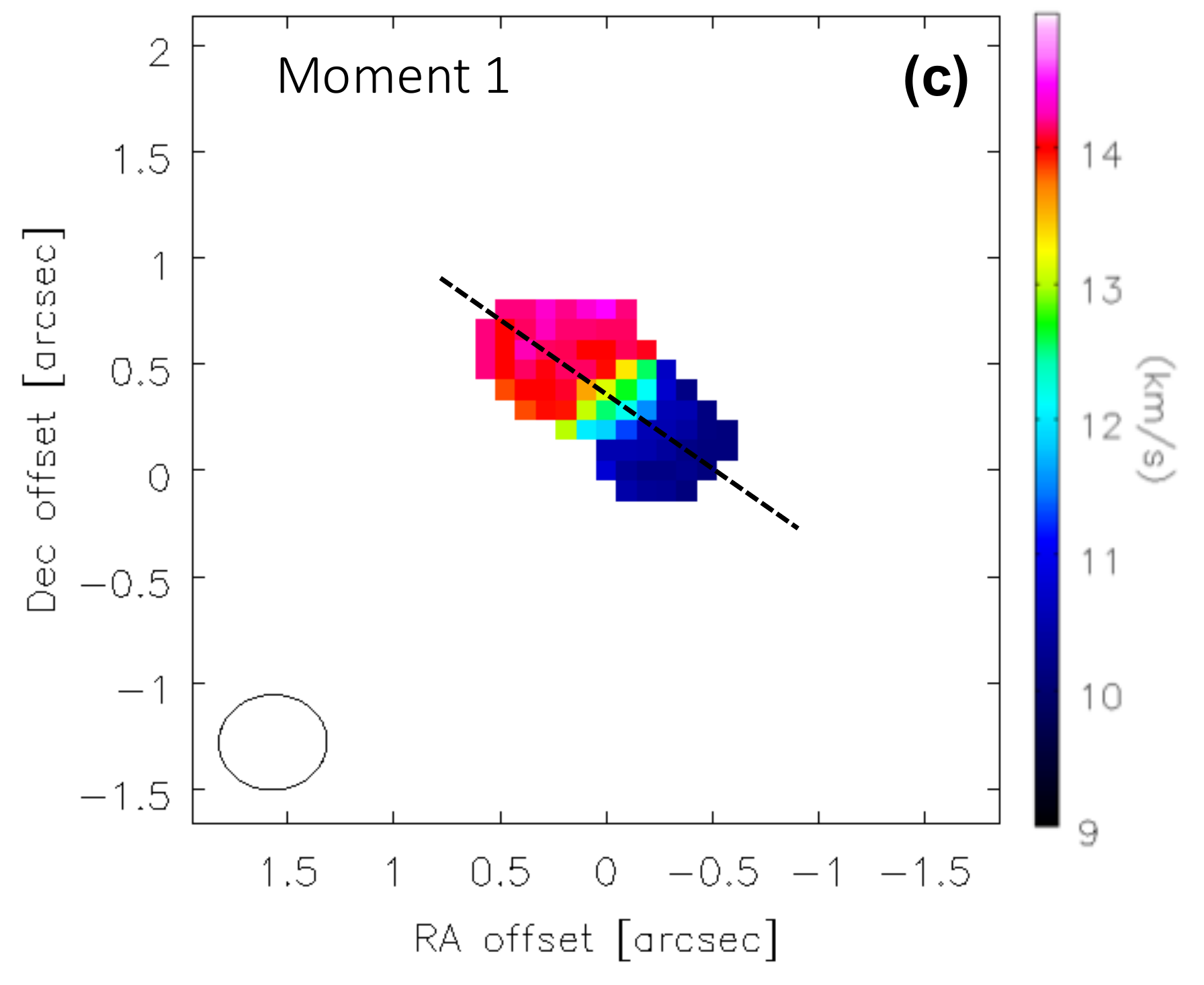}     
     \includegraphics[width=3.17in]{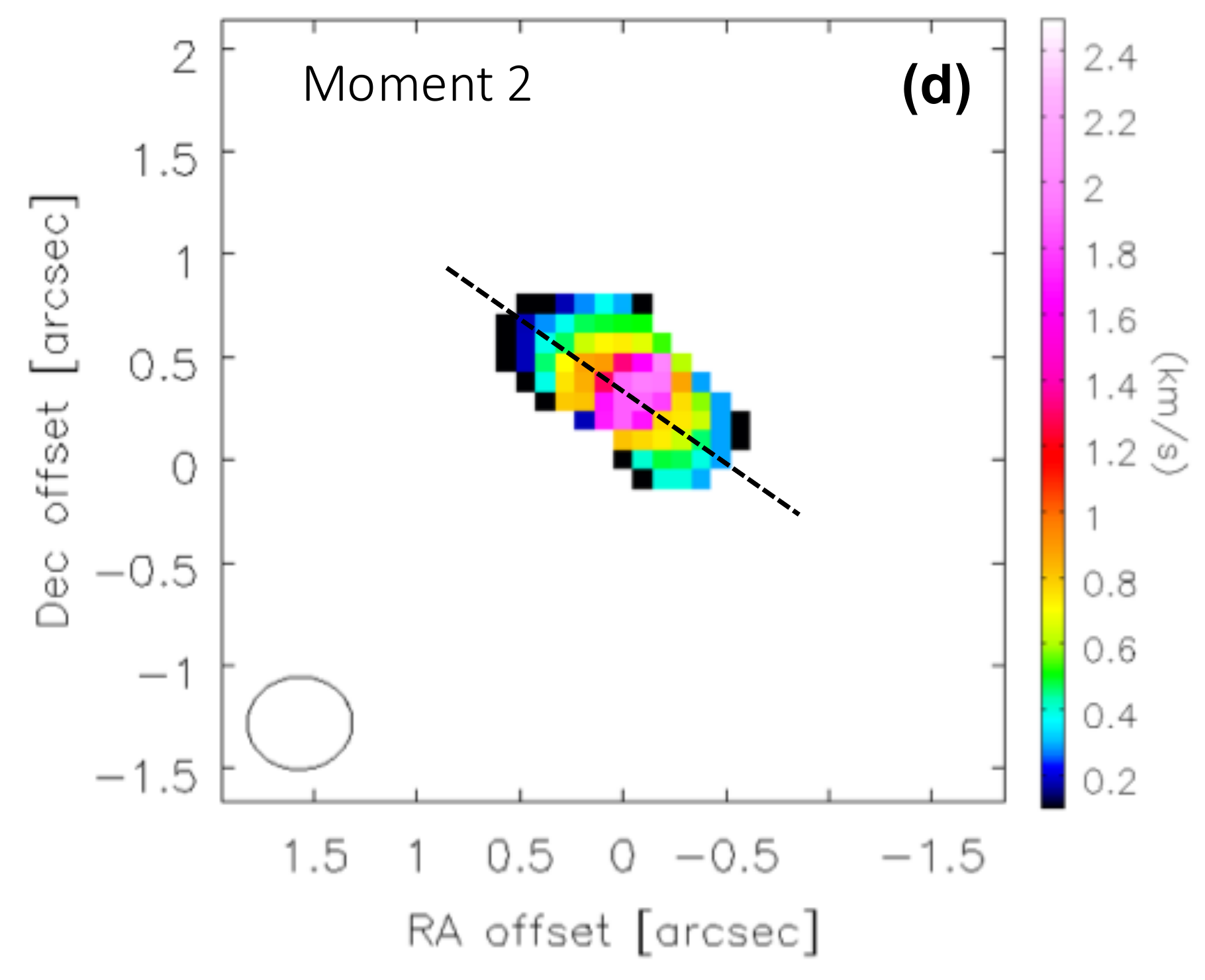}          
     \caption{The moment maps in the $^{12}$CO (2-1) line for M1701117. {\bf (a)} Moment 0 map. Red dashed line is along a PA of $\sim$45$\degr$. {\bf (b)} The 1.37 mm continuum map (black contours) overplotted on the moment 0 raster map. The contour levels are from 2-$\sigma$ to 10-$\sigma$ in steps of 2-$\sigma$. Red dashed line is the PA measured in the moment 0 map ($\sim$45$\degr$); white dashed line is the disc axis ($\sim$55$\degr$) measured in the continuum map. {\bf (c)} Moment 1 map and {\bf (d)} moment 2 map. Black dashed line is along the disc axis. The xy-axes show the position offset relative to the position of M1701117. North is up, east is to the left. }
     \label{moment}
  \end{figure*}

Figure~\ref{moment} shows the moment maps in the $^{12}$CO (2-1) line for M1701117. The moment 0 map shows the integrated CO emission elongated along a PA of $\sim$45$\degr$ (red dashed line in Fig.~\ref{moment}a), which is slightly offset by $\sim$10$\degr$ from the disc axis measured in the continuum map (Fig.~\ref{moment}b). The moment 1 map shows that the velocity gradient is along the disc axis (Fig.~\ref{moment}c), centered at a velocity of $\sim$12.5 km s$^{-1}$, and the red-shifted lobe is towards the north-east of the source. We can therefore consider the disc axis with a PA of 55.0$\degr \pm$5.6$\degr$ as the major axis along the velocity gradient, and a source V$_{LSR}\sim$ 12.5$\pm$0.5 km s$^{-1}$ for M1701117. As noted, it is also in the velocity channels of 12.5$\pm$0.7 km s$^{-1}$ that the peak in the observed CO emission is seen close to the source position (Fig.~\ref{ch-maps}). The moment 2 map shows an increase in the velocity dispersion towards the central proto-brown dwarf and emphasizes the faint high velocity emission near the central position, which is expected for Keplerian motion. It also shows a slight velocity dispersion along the jet direction towards the north-west, suggesting that there may also be some unresolved jet emission in the CO line.

%moments=0   - integrated value of the spectrum
%moments=1   - intensity weighted coordinate;traditionally used to get 'velocity fields'
%moments=2   - intensity weighted dispersion of the coordinate; traditionally used to get "velocity dispersion"

%Moment 0, shows the integrated CO emission elongated along the disc direction.
%Moment 1, shows the velocity gradient across the disk direction, centered on v=12.5km/s.
%Moment 2, shows the increase in velocity dispersion towards the protostar which is expected for Keplerian motion.

%It also shows increased velocity dispersion along the jet direction to the NW, so yes there is some unresolved jet emission too.

%PA=40deg, but it emphasizes the faint high velocity emission near the central position (expected for Keplerian emission).

%A similar elongation in the observed emission along the north-east to south-west direction is seen in the 1.3 mm continuum map (Fig.~\ref{cont}) and the CO line moment 0 map (Fig.~\ref{moment0}) for M1701117. 

%     The contours represent the signal-to-noise levels from 2-$\sigma$ to 10-$\sigma$ in steps of 2-$\sigma$. The 1-$\sigma$ rms is 0.8 mJy in the 1.37 mm continuum map.

\subsection{Position-Velocity Diagrams}
\label{PVDs}

\subsubsection{$^{12}$CO line}
\label{COlinePVDs}

\begin{figure*}
  \centering              
     \includegraphics[width=3.4in]{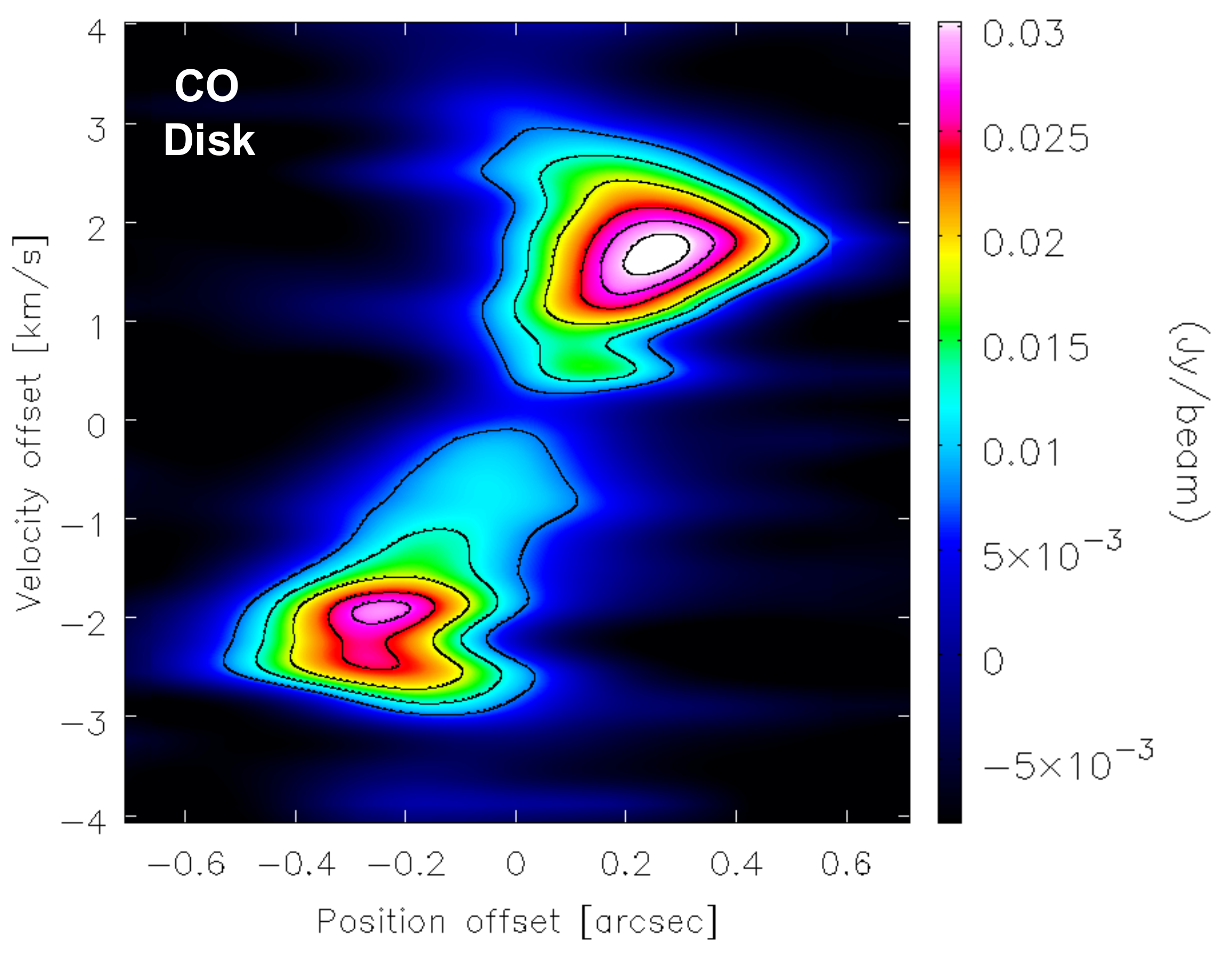} \hspace{0.2in}
         \includegraphics[width=3.2in]{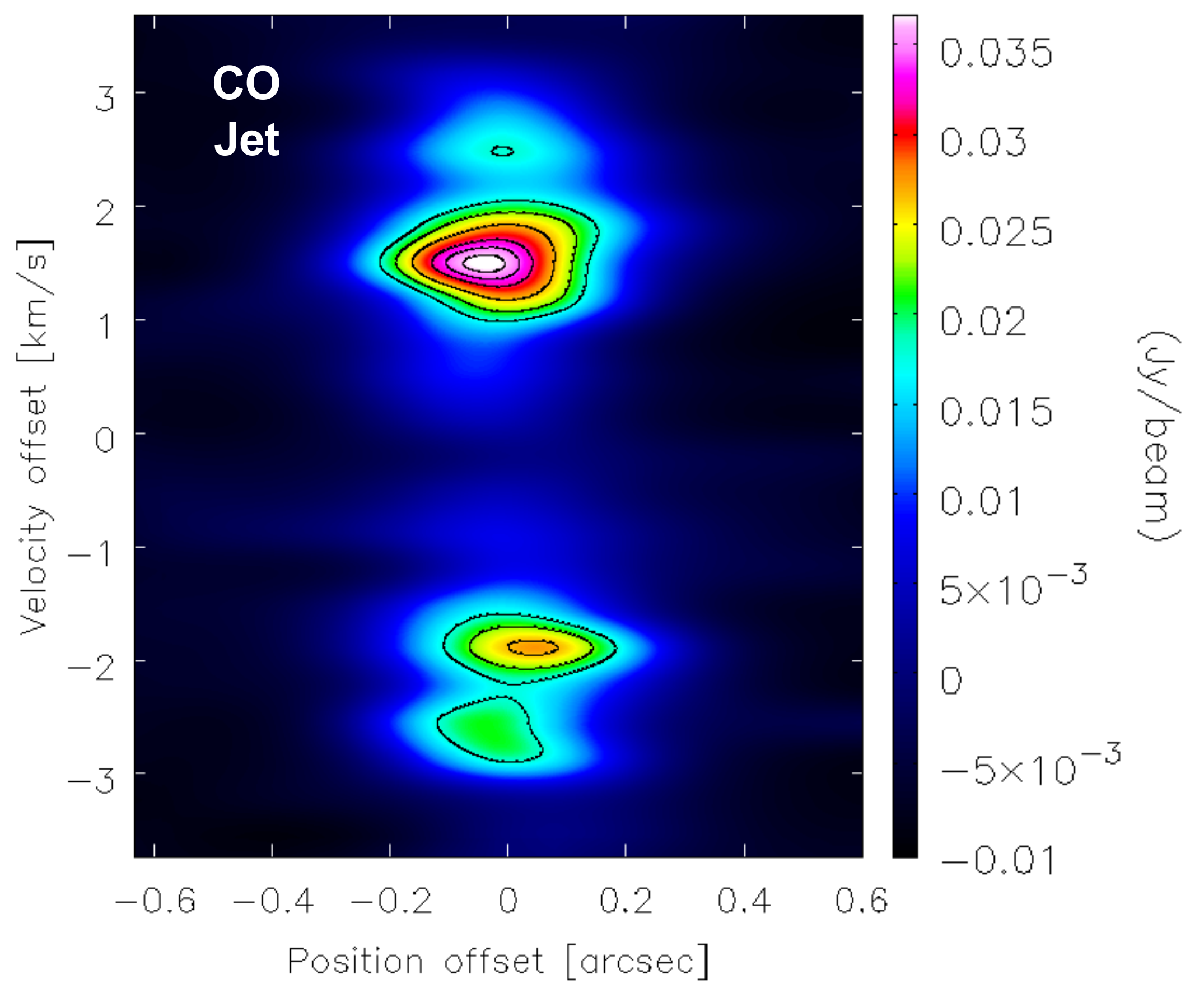} 
     \caption{The PV diagram in the CO line emission for a cut along the disc axis (left) and the jet axis (right). The velocity offset is with respect to the source V$_{LSR}$ of 12.5 km s$^{-1}$. The colour bar shows the flux scale in units of Jy beam$^{-1}$. The contours represent the signal-to-noise levels from 2-$\sigma$ to 10-$\sigma$ in steps of 2-$\sigma$. The 1-$\sigma$ rms is 11 mJy in the disc PVD and 14 mJy in the jet PVD. The red-shifted lobe is located towards the north-east of the source. }%North is up, East is to the right. }
     \label{pvd-co}
  \end{figure*}

The morphology of the observed line emission can be visualized more clearly in the image-space position-velocity diagram (PVD). Figure~\ref{pvd-co} shows the CO line PVDs created by making a cut along the disc axis (left panel) and the jet axis (right panel). For the disc PVD, the cut was made from the north-east to the south-west direction. For the jet PVD, the cut was made from the north-west to the south-east direction, which implies that the brighter red-shifted lobe is towards the north-west of the driving source (Fig.~\ref{maps}). The peak emission along the disc axis is seen at a (position, velocity) of (+0.26$\arcsec$, +1.7 km s$^{-1}$) and (-0.24$\arcsec$, -1.9 km s$^{-1}$). The peak emission along the jet axis is seen at a (position, velocity) of (-0.037$\arcsec$, +1.5 km s$^{-1}$) and (+0.034$\arcsec$, -1.9 km s$^{-1}$). The uncertainty in the (position, velocity) measurement is $\pm$0.05$\arcsec$ and $\pm$0.15 km s$^{-1}$. 

%The jet PVD shows the peak emission originating from close to the central source, and has revealed a compact, unresolved CO molecular outflow with a spatial extent of $<$0.4$\arcsec$ ($<$155 AU). 

%The jet PVD has revealed a CO molecular outflow associated with the HH~1165 jet. This compact, unresolved outflow component shows the peak CO emission originating from close to the central source, with a spatial extent of $<$0.4$\arcsec$ ($<$155 AU). 

The disc PVD shows a Keplerian like structure, with a spatial distance between the two emission peaks of $\sim$0.5$\arcsec$, about the same as the angular resolution of the observations. The jet PVD shows a compact (unresolved) CO molecular outflow, with peak emission originating from close to the central source and a spatial extent of $<$0.4$\arcsec$ ($<$155 AU). Interestingly, the CO jet PVD does not show any extended emission reaching up to the maximum recoverable scale of $\sim$4$\arcsec$; the molecular outflow is quite compact, unlike the much extended ($\sim$0.26 pc) atomic jet. Likewise, the disc PVD also does not show emission much extended beyond $\sim$0.5$\arcsec$, indicating that there is no extended infalling envelope in the outer part of the system. A comparison of the disc and jet PVDs (Fig.~\ref{pvd-co}) suggests that the disc emission seen at very low velocities close to the source V$_{LSR}$ must have some contamination from the molecular outflow, possibly originating from the boundary layers between the inner regions of the envelope/disc and the jet/outflow. This was also noted in the moment 2 map where an increasing velocity dispersion is seen along the jet direction towards the north-west (Fig.~\ref{moment}d).

A clear difference is seen in the structure of the red- and blue-shifted lobes along both axes. In the disc PVD, the red-shifted NE lobe appears flattened and elongated along the spatial direction, whereas the blue-shifted SW lobe shows elongation along the velocity scale. A similar elongation is seen in the blue-shifted SE lobe in the jet PVD (Fig.~\ref{pvd-co}). The elongation seen in the blue-shifted lobe could be due to an interaction of the outflow wind with the inner envelope/pseudo-disc regions. The optical spectra for M1701117 show emission in the K I~$\lambda\lambda$7665,7699 doublet and the $\sim$7300~\AA~[OII] line (Riaz et al. 2017), which are known to be kinematically associated with a high-velocity outflow wind (e.g., Hillenbrand et al. 2012). The M170117 system is viewed at a close to edge-on inclination of $\sim$60$\degr$-70$\degr$ (Sect.~\ref{COmodels}); inclination effects could be responsible for seeing the wind effects in just one lobe. There is also a brightness asymmetry in the CO line PVDs, with the northern red-shifted lobe being brighter than the southern lobe along both disc and jet axes. The asymmetry in morphology and brightness is similar to the one seen in the optical HH~1165 jet (Sect.~\ref{discussion}).

%It is not clear why the wind effects are only seen in one lobe. 

%The asymmetries can be explained by the disk fragmentation model. 

%diskContours: 0.0221 0.0241 0.0262 0.0282 0.0293 0.0301
%jetContours: 0.0273 0.0297 0.032 0.0344 0.0355 0.0365 

%DiskPVD Contours: 0.0218 0.0247 0.0277 0.0336 0.0366 0.0389  Jy/beam
%JetPVD Contours: 0.0264 0.029 0.0315 0.0341 0.0354 0.0364  Jy/beam

%Disk-Jet PVD: Contours: 0.0261 0.0271 0.0282 0.0292 0.0298 0.0302 Jy/beam for disk and Contours: 0.0273 0.0297 0.032 0.0344 0.0355 0.0365 Jy/beam for jet. the base contour level was set to 0.023 Jy/beam. 

\subsubsection{C$^{18}$O, H$_{2}$CO, N$_{2}$D$^{+}$ lines}
\label{bonus-lines}

\begin{figure}
  \centering              
     \includegraphics[width=3.1in]{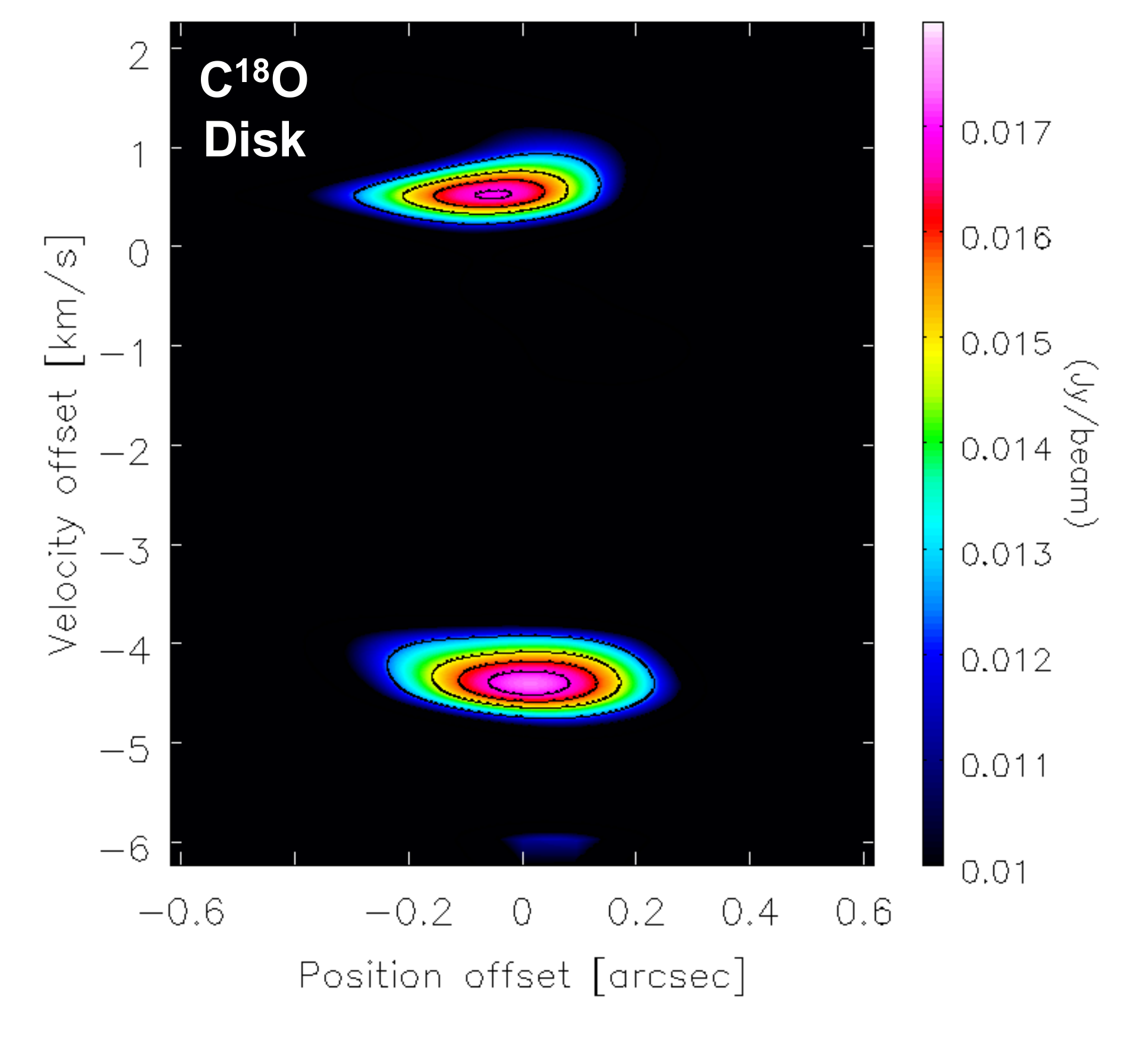} \hspace{0.2in}
     \includegraphics[width=3.1in]{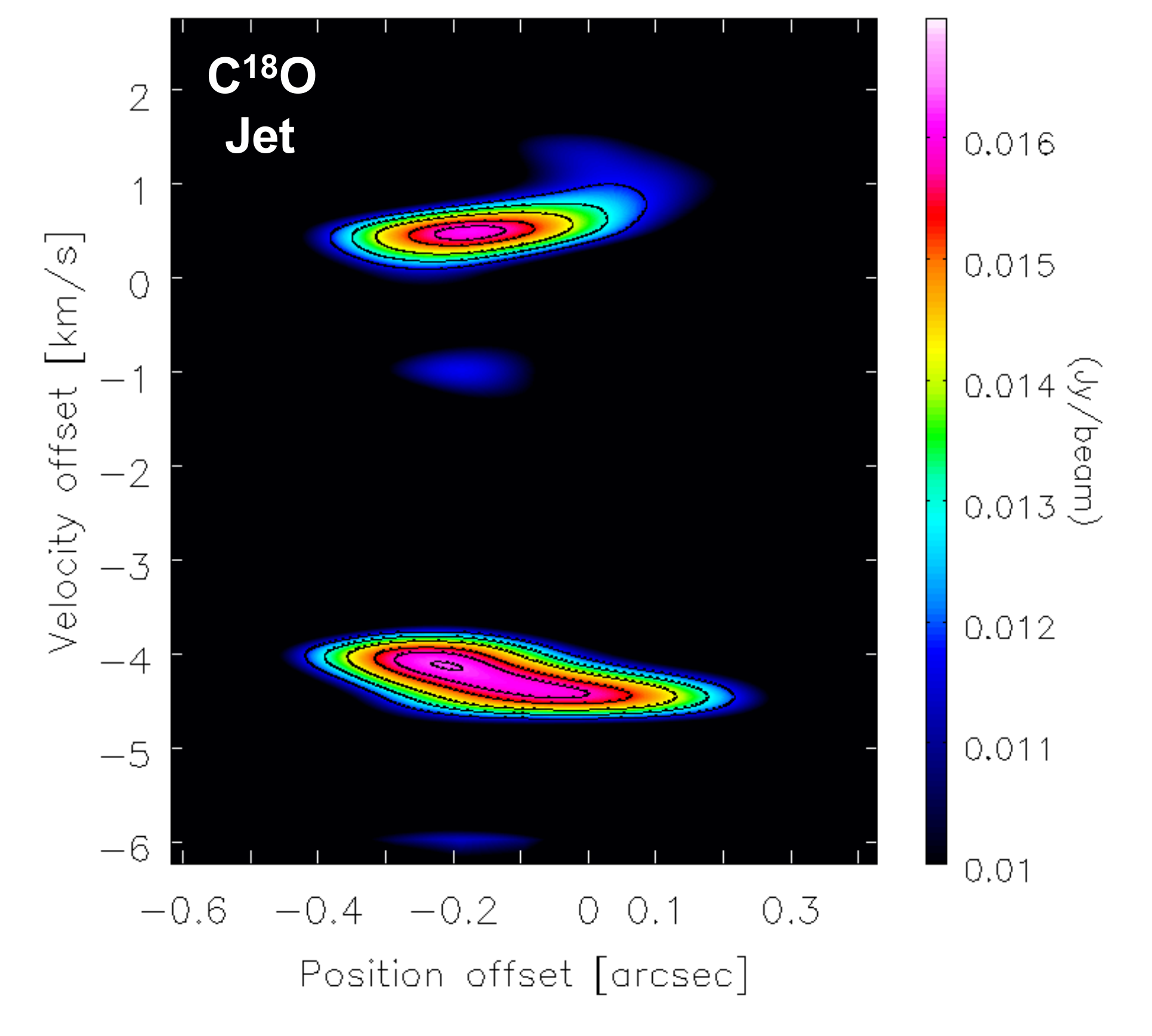}
     \caption{The PV diagrams in the C$^{18}$O line emission for a cut along the disc axis (left) and the jet axis (right). The velocity offset is with respect to the source V$_{LSR}$ of 12.1 km s$^{-1}$. The contour levels are 0.012, 0.013, 0.014, 0.015, 0.016, 0.017 Jy beam$^{-1}$. The base contour level is set to 0.011 Jy beam$^{-1}$ to enhance the peak emission positions. The red-shifted lobe is located towards the north-east of the source.   }
     \label{pvd-c18o}
  \end{figure}

\begin{figure*}
  \centering              
     \includegraphics[width=3.25in]{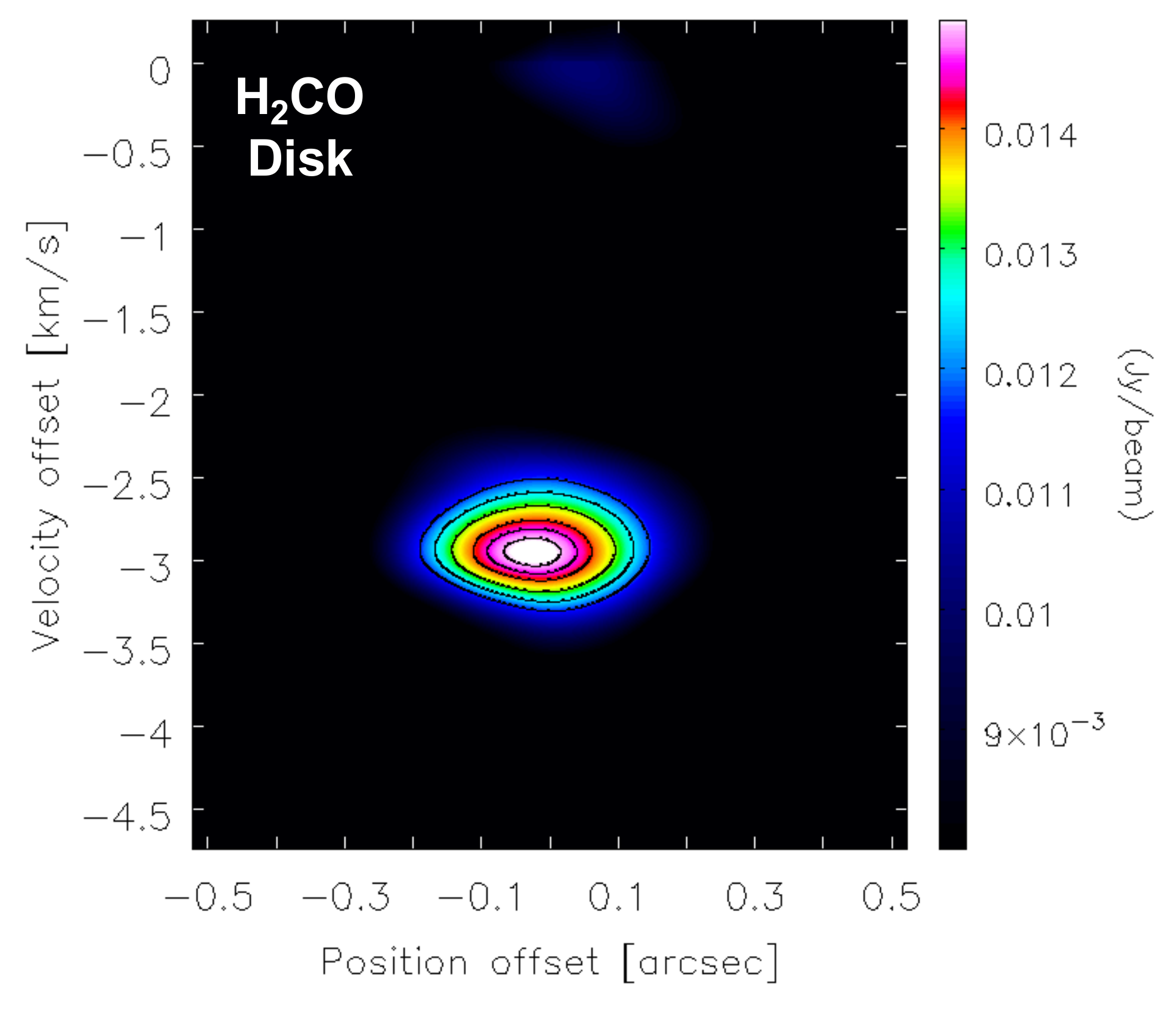} \hspace{0.2in}
     \includegraphics[width=3.1in]{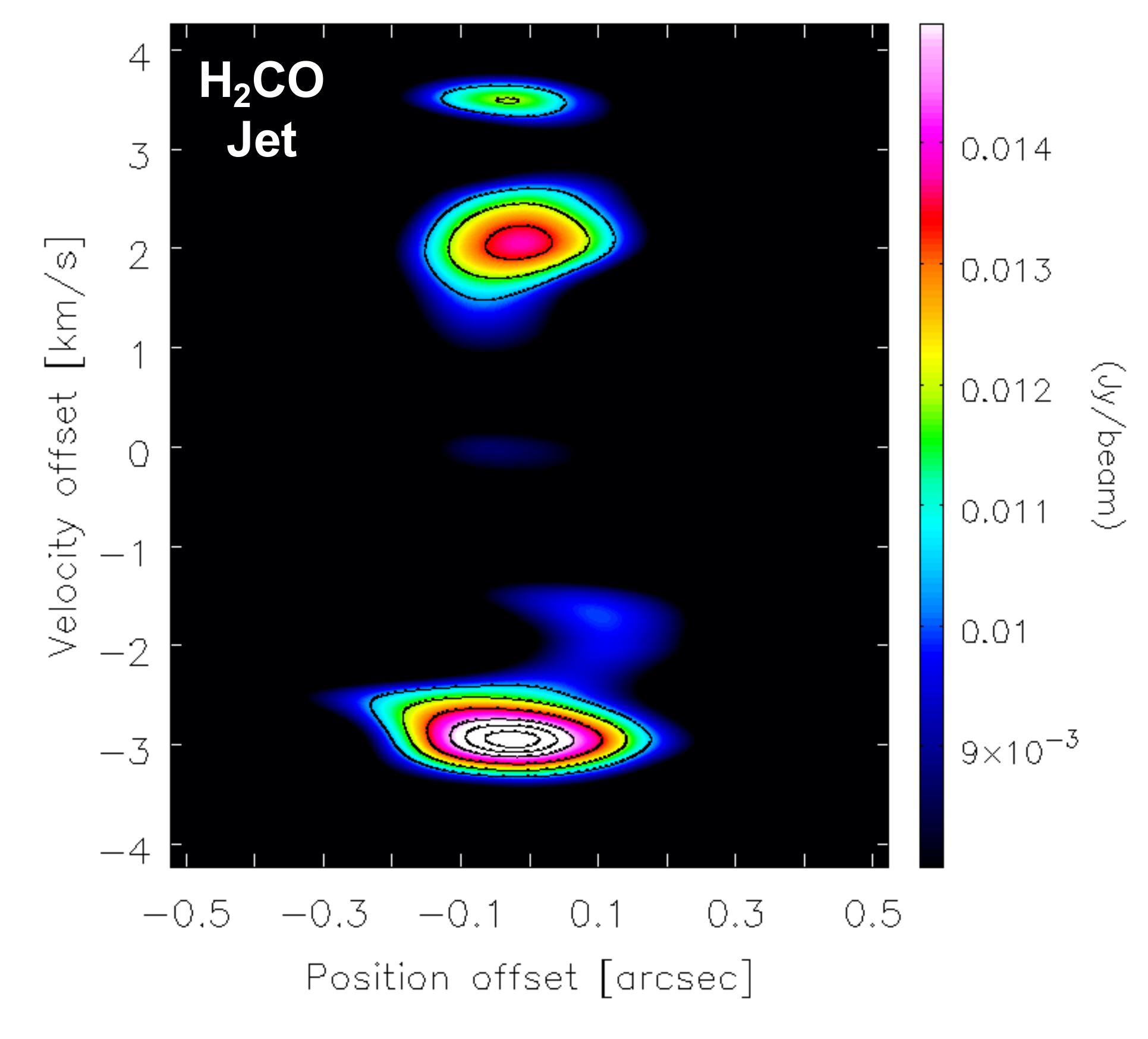}
     \caption{The PV diagrams in the H$_{2}$CO line emission for a cut along the disc axis (left) and the jet axis (right). The velocity offset is with respect to the source V$_{LSR}$ of 12.1 km s$^{-1}$. The contour levels are 0.01, 0.012, 0.013, 0.014, 0.015, and 0.016 Jy beam$^{-1}$. The base contour level is set to 0.01 Jy beam$^{-1}$ to enhance the peak emission positions. The blue-shifted lobe is located towards the south-west of the source.  }
     \label{pvd-h2co}
  \end{figure*}

\begin{figure}
  \centering              
     \includegraphics[width=3in]{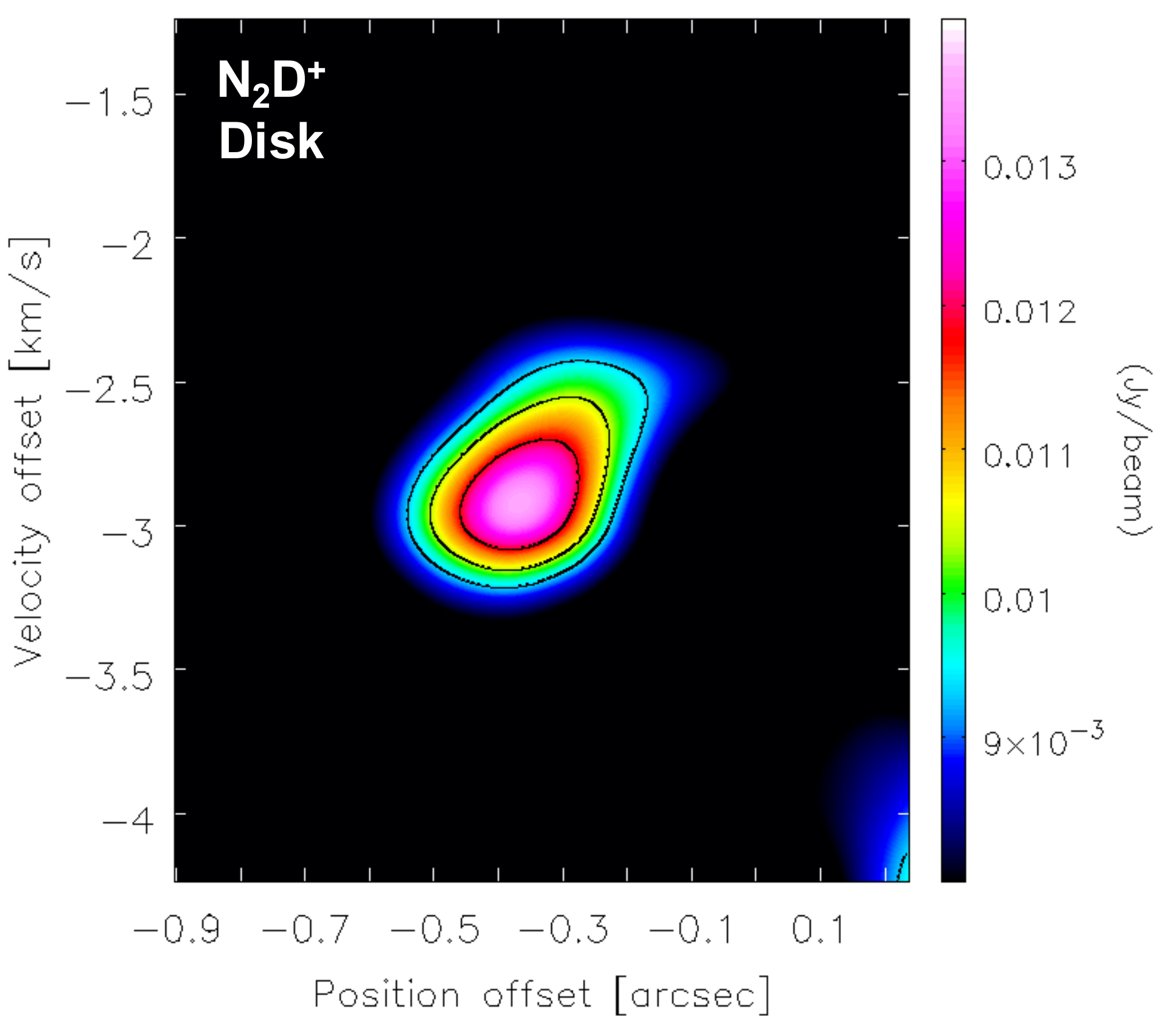}
     \caption{The PV diagrams in the N$_{2}$D$^{+}$ line emission for a cut along the disc axis. The velocity offset is with respect to the source V$_{LSR}$ of 12.1 km s$^{-1}$. The contour levels are 0.01, 0.012, 0.013, 0.014, 0.015 Jy beam$^{-1}$. The base contour level is set to 0.01 Jy beam$^{-1}$ to enhance the peak emission positions. The blue-shifted lobe is located towards the south-west of the source.   }
     \label{pvd-n2dp}
  \end{figure}

Figures~\ref{pvd-c18o},~\ref{pvd-h2co}, and \ref{pvd-n2dp} show the PVDs in the C$^{18}$O (2-1), H$_{2}$CO (3-2), and N$_{2}$D$^{+}$ (3-2) lines. Note that these are weak line detections with peak intensities a factor of $\sim$2 lower than CO. C$^{18}$O and H$_{2}$CO emission is detected along both disc and jet axes, while N$_{2}$D$^{+}$ shows a weak detection only along the disc axis. C$^{18}$O disc PVD shows peak emission close to the driving source at the (position, velocity) offset of (-0.08$\arcsec$, +0.5 km s$^{-1}$) and (0.06$\arcsec$, -4.1 km s$^{-1}$), while the jet PVD shows peaks at the same velocity offset but shifted towards the south-east at an offset position of -0.2$\arcsec$. The C$^{18}$O brightness is similar in the blue- and red-shifted lobes in both the disc and jet PVDs (Fig.~\ref{pvd-c18o}). C$^{18}$O emission is un-associated with the disc or jet emission in CO, as noted by the different position and velocity offsets, and the emission likely arises from the surrounding cloud material.

H$_{2}$CO emission is seen close to the driving source, with a spatial extent of $<$0.2$\arcsec$ (Fig.~\ref{pvd-h2co}). The peak H$_{2}$CO emission along the disc axis is at a (position, velocity) of (0$\arcsec$, -3 km s$^{-1}$), and along the jet axis at (0$\arcsec$, +2 km s$^{-1}$) and (-0.02$\arcsec$, -3 km s$^{-1}$). The blue-shifted emission in the jet PVD is brighter than the red-shifted lobe. The position and velocity offsets in H$_{2}$CO along the disc axis are different than observed in CO; the H$_{2}$CO emission arises from close to the central proto-brown dwarf and at a comparatively larger velocity offset than in CO. An exception is the H$_{2}$CO jet PVD where the red-shifted peak coincides with the red-shifted lobe in CO (Figs.~\ref{pvd-h2co};~\ref{pvd-co}), indicating an association of the H$_{2}$CO emission with the CO outflow.

The N$_{2}$D$^{+}$ disc PVD shows a single peak at a (position, velocity) of (-0.36$\arcsec$, -2.9 km s$^{-1}$) (Fig.~\ref{pvd-n2dp}). Both N$_{2}$D$^{+}$ and H$_{2}$CO emission show a similar velocity offset along the disc axis. A comparison with the blue-shifted lobe in the CO disc PVD (Figs.~\ref{pvd-n2dp};~\ref{pvd-co}) shows a similar position but a comparatively larger velocity offset than in CO. N$_{2}$D$^{+}$ emission is more compact compared to the extended blue-shifted lobe in CO, and the peak intensity in N$_{2}$D$^{+}$ is a factor of $\sim$2.6 lower than CO. The possible reasons for only detecting the blue-shifted emission in N$_{2}$D$^{+}$ and H$_{2}$CO are discussed in Sect.~\ref{discussion}.

%possibly caused by external irradiation effects (Sect.~\ref{photo-ev}). 

%The elongation seen in the blue-shifted lobe could be due to an interaction of the outflow wind with the inner envelope/pseudo-disk regions.

%But n2dp shows a peak at the very center and has survivved in the gas phase where both co and dcop are depleted. 

%The velocity offset in N$_{2}$D$^{+}$ is the same as in the H$_{2}$CO disk PVD, and both species have been detected in cold, dense proto-brown dwarfs (Riaz et al. 2018 a,b). 

%perhaps the n2dp position offset is due to disk fragmentation. the edge may be fragmented. either new clump being formed or it may be a high density clump depleted in co but has n2dp emission. then cite lis paper. 

%perhaps the n2dp disk offset is due to fragmentation. the edge may be fragmented. either new clump being formed or it may be a high density clump depleted in co but has n2dp emission. then cite lis paper. 

%As noted, the peak position in C$^{18}$O and H$_{2}$CO is close to the source at 0$\arcsec$, while the velocity offsets are further blue-shifted to $\sim$3-4.5 km s$^{-1}$. 

%only c18o is completely offset. N2dp is slightly offset but w overlap w disk co, h2co overlap w jet co. no ovelap in h2co disk. 

%H2CO Disk Contours: 0.0117 0.0125 0.0133 0.0141 0.0145 0.0148 Jy/beam
%H2CO Jet Contours: 0.0104 0.0118 0.0133 0.0147 0.0155 0.0161

%N2Dp Contours: 0.0104 0.0116 0.0127 0.0139 0.0145 0.015

%C18O Contours: 0.0141 0.0151 0.0162 0.0172 0.0177 0.0181
%C18O Disk Contours: 0.00766 0.01 0.0124 0.0147 0.0159 0.0169
%C18O Jet Contours: 0.0118 0.0129 0.0141 0.0152 0.0158 0.0163

\section{Comparison with Models}
\label{model-comp}

\subsection{Source Properties}
\label{source}

\begin{figure}
\center 
     \includegraphics[width=80mm]{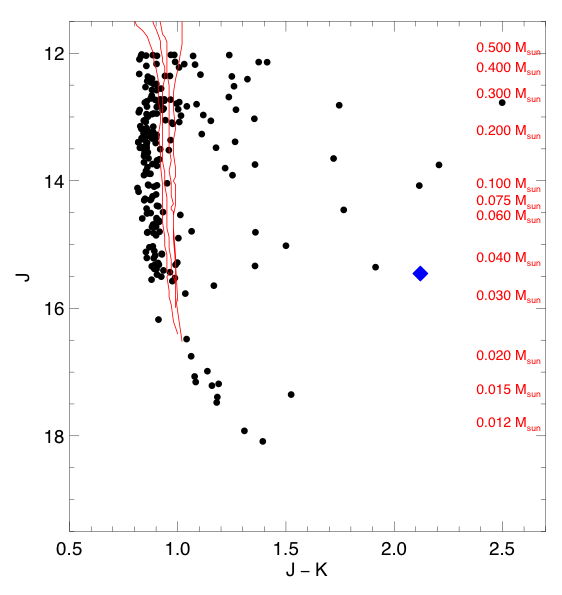} 
    \caption{The UKIDSS cmd with the known $\sigma$ Orionis cluster members marked as filled circles, and M1701117 marked as a blue triangle. The red lines are pre-main sequence isochrones from the Baraffe et al. (2003) models, and are at ages of 1, 3, 5, and 8 Myr ({\it right to left}). The mass scale shown on the right hand side is from the 1 Myr model isochrone. } 
    \label{ukidss-cmd} 
\end{figure}

We revisit the total (star + circumstellar) mass estimate for the M1701117 system, and the likelihood of it evolving into a brown dwarf. From ALMA continuum observations, the total (dust+gas) mass arising from the circumstellar material in the M1701117 system is 20.98$\pm$1.24 M$_{Jup}$. To estimate the intrinsic stellar mass for M1701117, we have used the infrared photometry for this object and the evolutionary models by Baraffe et al. (2003). The $J$-band, in particular, is considered to be least affected by the potential effects of veiling at bluer wavelengths, and circumstellar disc emission further into the infrared (e.g., Hartigan et al. 1995; White \& Hillenbrand 2004; Cieza et al. 2005). Figure~\ref{ukidss-cmd} shows the UKIDSS ($J-K, J$) colour-magnitude diagram (cmd). The filled circles represent known members from previous studies in the $\sigma$ Orionis cluster (Bejar et al. 1999, 2001; Caballero et al. 2007; Caballero 2008; Zapatero Osorio et al. 2000). The cluster suffers from negligible reddening (A$_{V}$$\leq$1 mag; B\'{e}jar et al. 1999) and the photometric data has been corrected for the interstellar reddening. M1701117 appears much redder than the main cluster member sequence, which is due to the large $K$-band excess emission from the circumstellar material. We have over plotted in Fig.~\ref{ukidss-cmd} the 1, 3, 5, and 8 Myr isochrones from the Baraffe et al. (2003) models. The isochrones at ages of $<$1 Myr are not provided in these models. The mass scale shown on the right hand side in Fig.~\ref{ukidss-cmd} is from the 1 Myr model isochrone at a distance of 410 pc for the cluster. Assuming an age of 1 Myr, we can estimate a stellar mass of $\sim$40 $M_{Jup}$ and a stellar luminosity of 0.012 L$_{\sun}$ for M1701117. The location of M1701117 to the red of the 1 Myr isochrone in the cmd indicates an age younger than 1 Myr. The typical age of the embedded phase or the age of the protostar formation is considered to be $\sim$0.01-0.05 Myr, based on the statistics of embedded-to-Class II sources (e.g., Evans et al. 2009). On the other hand, the embedded phase may be shorter for very low mass protostars and brown dwarfs, as indicated by numerical simulations. The estimates on the stellar mass and luminosity for M1701117 will be lower if the object is younger than 1 Myr.

\begin{figure}
  \centering              
     \includegraphics[width=2.8in]{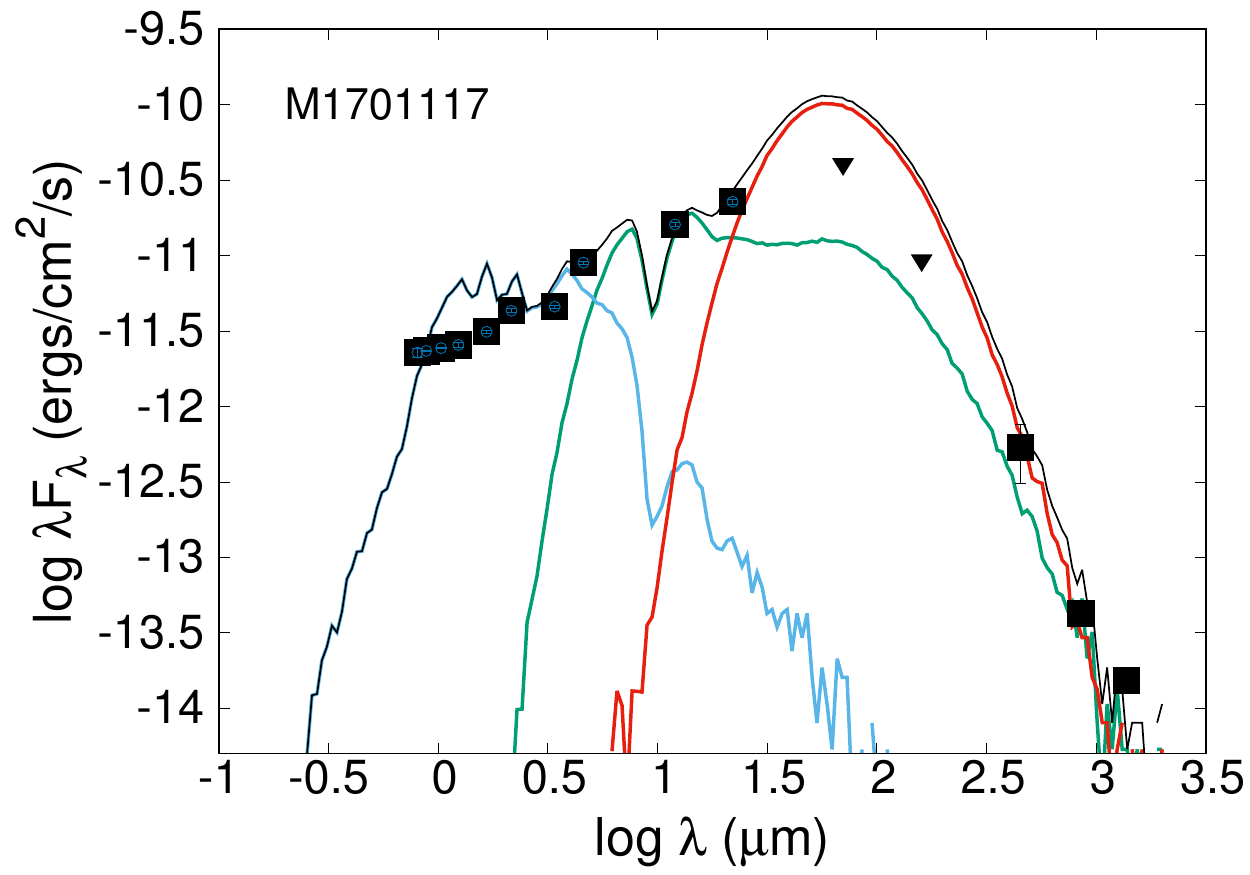}
     \caption{The SED with model fit for M1701117. The red, green, and blue lines indicate the individual contribution from the envelope, disc, and stellar components, respectively.  }
     \label{SED}
  \end{figure}

Figure~\ref{SED} shows the spectral energy distribution (SED) for M1701117 that now includes the new ALMA 1.37 mm data point. We refer to Riaz et al. (2015) for a detailed discussion on the continuum radiative transfer modeling and the related degeneracies in the model fit. The bolometric luminosity ($L_{bol}$) of the system, as obtained by integrating over the observed SED from optical to millimeter wavelengths, is 0.09$\pm$0.03 L$_{\sun}$. The Herschel PACS points are upper limits and suggest that the integrated intensity and thus the bolometric luminosity may be over-estimated. The 2-22$\mu$m slope of the SED for M1701117 is 0.67$\pm$0.02, higher than the threshold of 0.3 considered between Class 0/I and Class II objects. We can further use the ``Stage'' classification scheme to determine if this is a Class 0 or Class I object. In this scheme, Stage 0 objects have $M_{disc}$/$M_{env}$ $<<$ 1 and $M_{circum}$/$M_{star}$ $\sim$ 1 ($M_{circum}$ = $M_{disc}$ + $M_{env}$), while Stage I objects have 0.1 $<$ $M_{disc}$/$M_{env}$ $<$ 2 and $M_{circum}$ $<$ $M_{star}$ (Whitney et al. 2003; Robitaille et al. 2006). It is difficult to distinguish between the masses of the individual components. For M1701117, the mass of the pseudo-disc, $M_{circum}$, is $\sim$0.02 M$_{\sun}$, which implies $M_{circum}$/$M_{star} \sim$0.5. In Riaz et al. (2015), we estimated the ratio $M_{disc}$/$M_{env} \sim$0.002 from the best model fit to the SED. These ratios place M1701117 intermediate between Stage 0 and Stage I.

The spectral slope of the SED from 850 $\mu$m to 1.37 mm, $\alpha$, is related to the slope, $\beta$, of the opacity law, $\kappa \propto \nu^{\beta}$, as $\alpha \approx \beta$ + 2 in the Rayleigh-Jeans limit (e.g., Beckwith \& Sargent 1991). The spectral slope $\alpha$ for M1701117 is 2.1, which is consistent with the mean value of 2.4 measured for low-mass Class 0/I protostars (e.g., Joergensen et al. 2007). The optically thin envelope emission would have a spectral slope of 3.5-4 for typical dust opacities (e.g., Joergensen et al. 2007). Thus $\alpha \sim$2 for M1701117 suggests that the compact continuum emission is marginally optically thick, and has an origin both in the inner Keplerian disc and the circumstellar envelope at the very shortest baselines. The typical value of $\beta$ for dust in the optically thin interstellar medium is $\sim$1.7 (e.g., Lin et al. 2016), while grain growth up to millimeter sizes will result in $\beta <$1 (e.g., Draine 2006). An $\alpha$ of 2.1 for M1701117 implies $\beta \sim$0.1, indicating that significant grain growth has already occurred in the young pseudo-disc in this proto-brown dwarf. As a comparison, low-mass Class 0/I protostars show a wide range in $\beta$ between 0.3 and 2.0 (e.g., Li et al. 2017; Joergensen et al. 2007), while in the more evolved T Tauri and Herbig Ae/Be discs, the values of $\beta$ are in the range of 0.5-1.0 (e.g., Lommen et al. 2007; Andrews \& Williams 2005). Therefore, a range of grain growth levels is observed at any given stage with no evidence of a significant build up of discs from the Class 0 to the T Tauri stage. On the other hand, the very small observed values of $\beta$ may be interpreted by the presence of deeply embedded and hot inner discs, which only significantly contribute to the observed fluxes at long wavelength bands.

%Is it Class I or Class 0? 
%alpha = 0.67 +/- 0.02 (>0.3 is class 0/I)

%Discuss stage based on circum/star mass:

%stage 0 --> Mdisk/Menv << 1, Mcircum/Mstar ~ 1
%stage I --> 0.1 < Mdisk/Menv < 2 , Mcircum < Mstar

%For M170 SED fit: Mdisk = 0.1-1.5 MJ, Menv=30-65 MJ, Best values: Mdisk= 0.1 MJ, Menv=50 MJ ==> Mdisk/Menv = 0.002, Mcircum = Mdisk+Menv = 50.1 MJ

%From cont obs: Mcircum = 20 MJ, Mcircum/Mstar = 20/40 = 0.5, or Mcircum/Mstar = 20/60 =0.3
%the observed Mcircum is closer to the lower estimate on Menv=30 MJ. 

%while we cannot distinguish between the disk and envelope mass from the observed value, typically disk mass is 1-10\% of the env mass, consistent w what we find for m170. 

%450mu - 1.37mm slope -- beta would tell optically thin or not

%The use of isochrones for non-accreting models usually make the objects look older than they are; however, even with this artificial ageing, our object still appears quite young ($<$1 Myr). We have therefore considered the estimates on stellar parameters obtained using the non-accreting evolutionary models as good estimates on the intrinsic properties of the source. 
 
From the analysis above, a stellar mass of $\sim$0.04 M$_{\sun}$ and a circumstellar mass of $\sim$0.02 M$_{\sun}$ results in a total mass of $\sim$0.06 M$_{\sun}$ for the M1701117 system in the present epoch, which is below the sub-stellar limit ($\sim$0.075-0.08 M$_{\sun}$; Baraffe et al. 2003). Considering that this is a Stage 0/I object still accreting from its circumstellar material, this is the present but not the final mass estimate. The fact that we see signatures of an infalling pseudo-disc (Sect.~\ref{core}) and that the total mass of the circumstellar material is very low indicates that we are looking at the late accretion phase of star formation when the infalling material begins to dissipate. Being the driving source of HH~1165, $\sim$30-80\% of the infalling material in M1701117 is expected to be expelled by the jet (Machida et al. 2009). A comparison of the observed velocity spread with various models (Sect.~\ref{core}) also suggests this system to be more evolved than the dynamical age of the jet ($\sim$2900 yr; Riaz et al. 2017). This object is located in an irradiated environment (Riaz et al. 2017), which could further expedite the dissipation of the circumstellar material. Considering these scenarios, M1701117 is likely to reach a final mass below the sub-stellar limit and become a brown dwarf.

\subsection{CO Line Modeling}
\label{COmodels}

\begin{figure}
\center 
     \includegraphics[width=70mm]{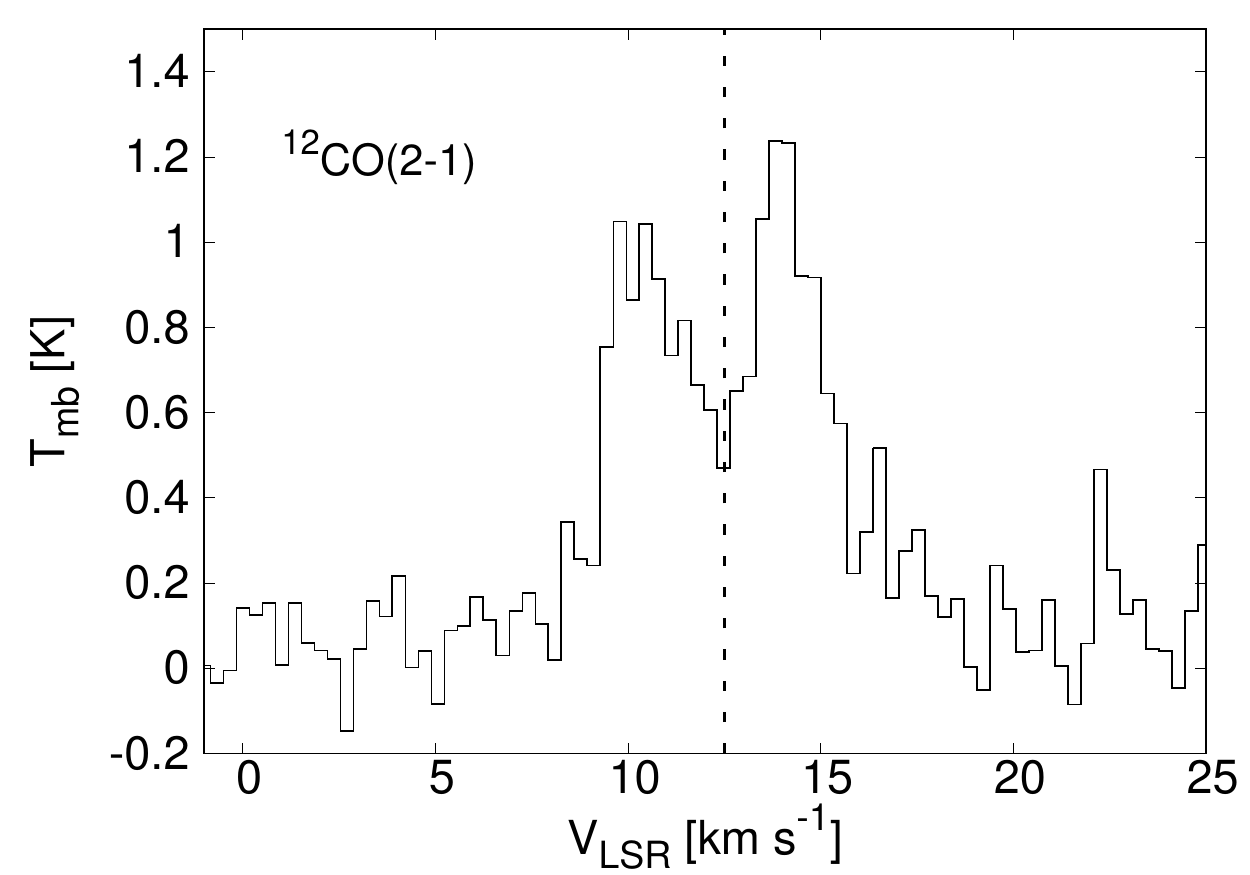} 
    \caption{The $^{12}$CO (2-1) spectrum for M1701117. Dashed line marks the source V$_{LSR}$. } 
    \label{co-line} 
\end{figure}

Since the brightest emission is seen in CO, this line is used for further study of the infall and rotational kinematics. Figure~\ref{co-line} shows the observed CO spectrum for M1701117 collapsed over all velocity channels. The double-peaked line profile with red-dominated asymmetry is reminiscent of Keplerian rotation as typically observed in disc sources. The line profile is mirrored at roughly about the source V$_{LSR}\sim$ 12.5$\pm$0.5 km s$^{-1}$. The CO line profile also shows broad, extended wings that are indicative of a contribution from a molecular outflow.

In order to interpret the observed morphology, we have used a coupled physical and chemical 3D radiative transfer model that can simultaneously provide a good fit to the CO spectrum, the velocity spread of $\pm$2 km s$^{-1}$, and the spatial spread of $\pm$77-96 AU, as seen in the CO disc PVD. We have employed various physical models with different structures, as discussed in Sect.~\ref{core},~\ref{fragment}, and ~\ref{kepler}. The physical structure, i.e., the density, temperature, and velocity radial profiles for a given model is used as an input to the 3D non-LTE radiative transfer code MOLLIE (Keto et al. 2004).  We generate synthetic CO spectra for different radial profiles of the molecular abundance. The synthetic CO line profiles are convolved by the ALMA 0.4$^{\prime\prime}$ beam size and compared with the observed CO spectrum for M1701117. A reasonable fit to the strength and width of the observed profile is reached from varying the abundance profile and the line width. For each synthetic spectrum, a reduced-$\chi^{2}$ value is computed to determine the goodness of fit. From the best line model fit (lowest reduced-$\chi^{2}$ value) to the observed CO profile, we produced a PVD for a cut along the disc axis (55$\degr$), and compared the position and velocity offsets in the model and observed PVDs.

%The line modeling method is the same for all models. 

%the problem is not the elongation or not a perfectly Keplerian profile, but the high velocity shift. 

%We observe a velocity spread of about $\pm$2 km s$^{-1}$, and a spatial spread of about $\pm$77-96 AU.

%Figures~\ref{co-line} shows the observed CO spectrum for M1701117 extracted along the disk axis and collapsed over all velocity channels. 

\subsubsection{Core Collapse Model}
\label{core}

We consider the physical structure from the core collapse model (hereafter; CC model) of brown dwarf formation presented in Machida et al. (2009). These are 3D magneto-hydro-dynamic (MHD) simulations of brown dwarf formation via gravitational collapse of a very low-mass core. The initial setup of the model, the numerical method and numerical settings are the same as those described in Machida et al. (2008; 2009; 2014). Using a 3D resistive MHD nested grid code and a barotropic equation of state, the evolution of the cloud is calculated from the pre-stellar cloud core stage until about 0.1 Myr after the proto-brown dwarf formation, in which about eight orders of magnitude of spatial scale (5 $\times$ 10$^{5}$ - 0.005 AU) is covered.

 \begin{figure}
  \centering     
     \includegraphics[width=2.7in]{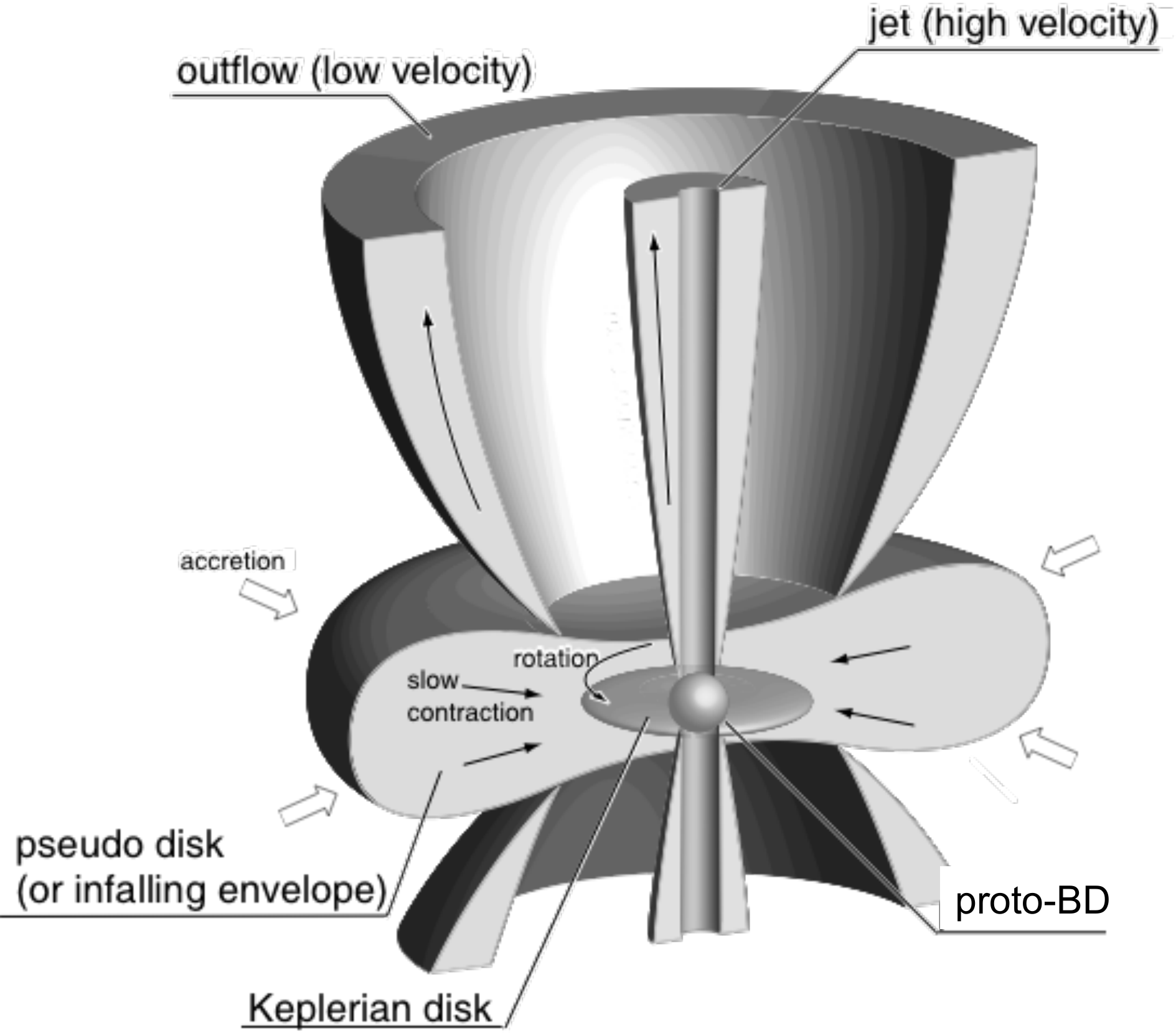}   
     \caption{Schematic view of the basic structure of circumstellar pseudo-disc, inner Keplerian disc, low-velocity outflow, and high-velocity jet from MHD simulations of brown dwarf formation via core collapse.   }
     \label{CC-model-1}
  \end{figure}

\begin{figure*}
  \centering     
     \includegraphics[width=2.7in]{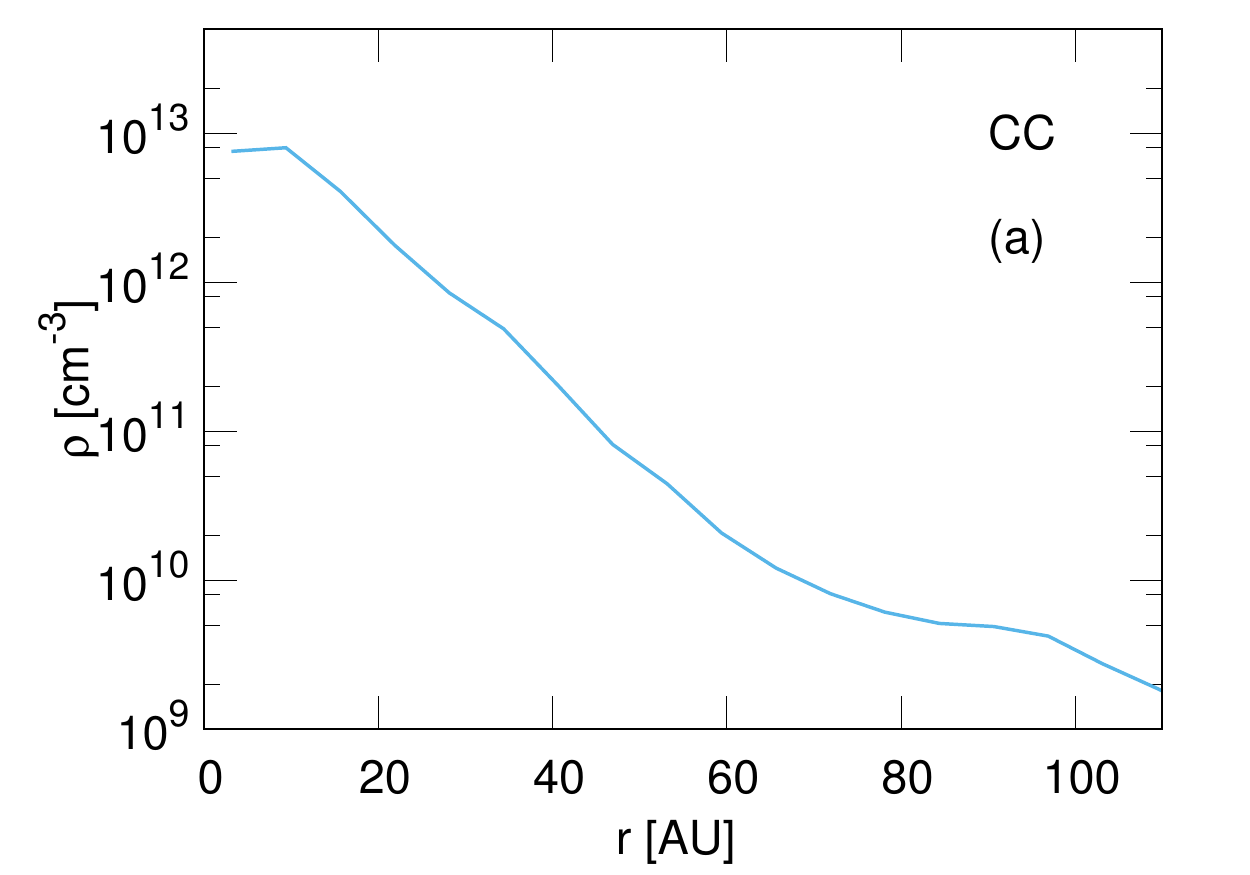}  
     \includegraphics[width=2.7in]{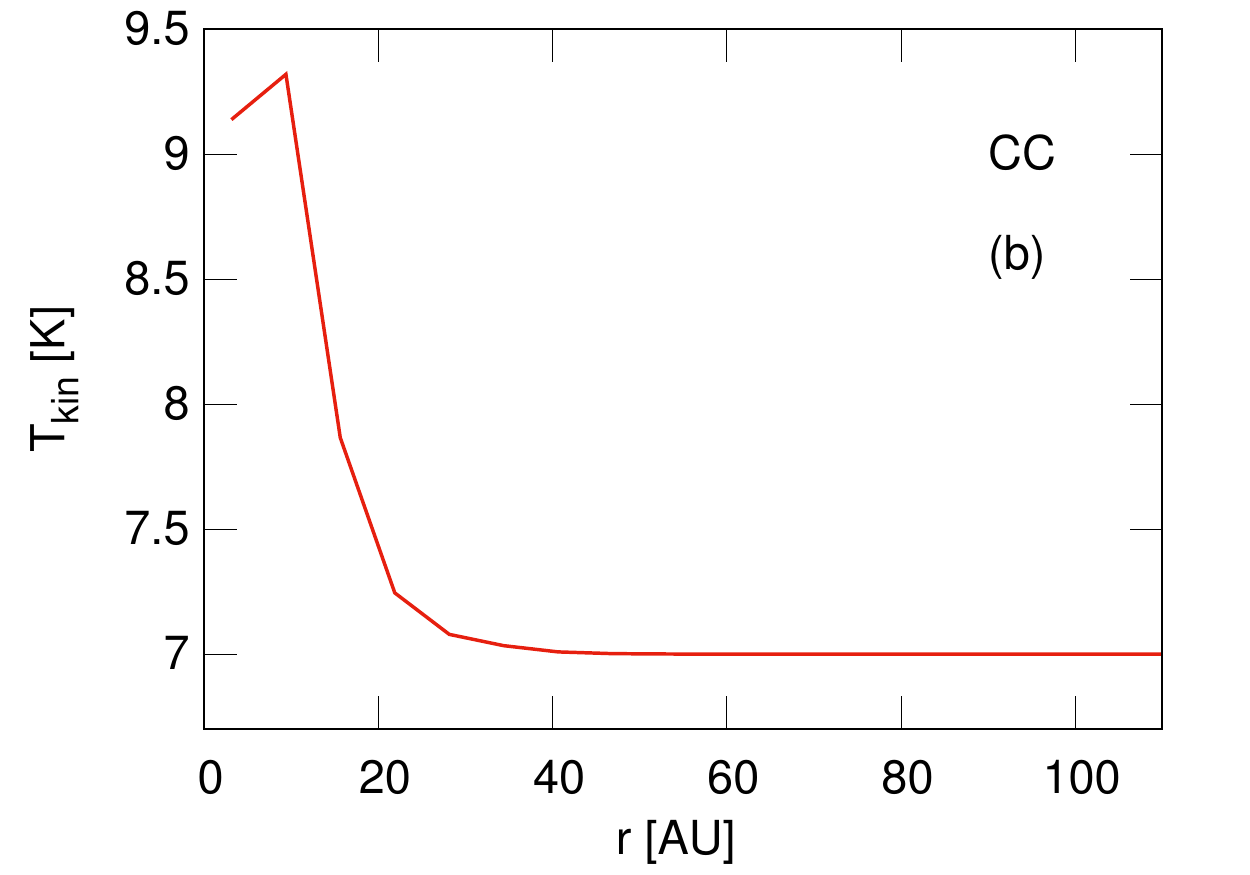}  
      \includegraphics[width=2.7in]{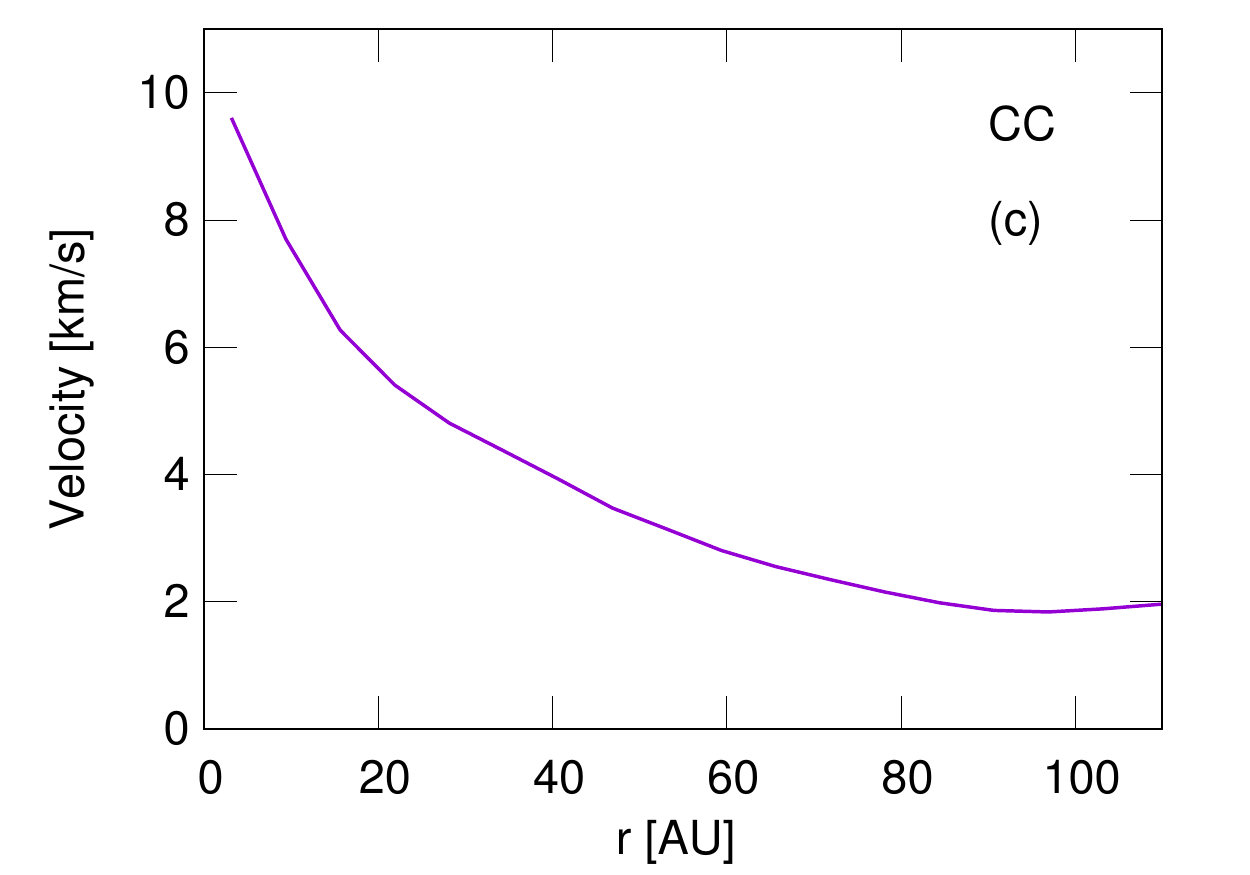}     
      \includegraphics[width=2.7in]{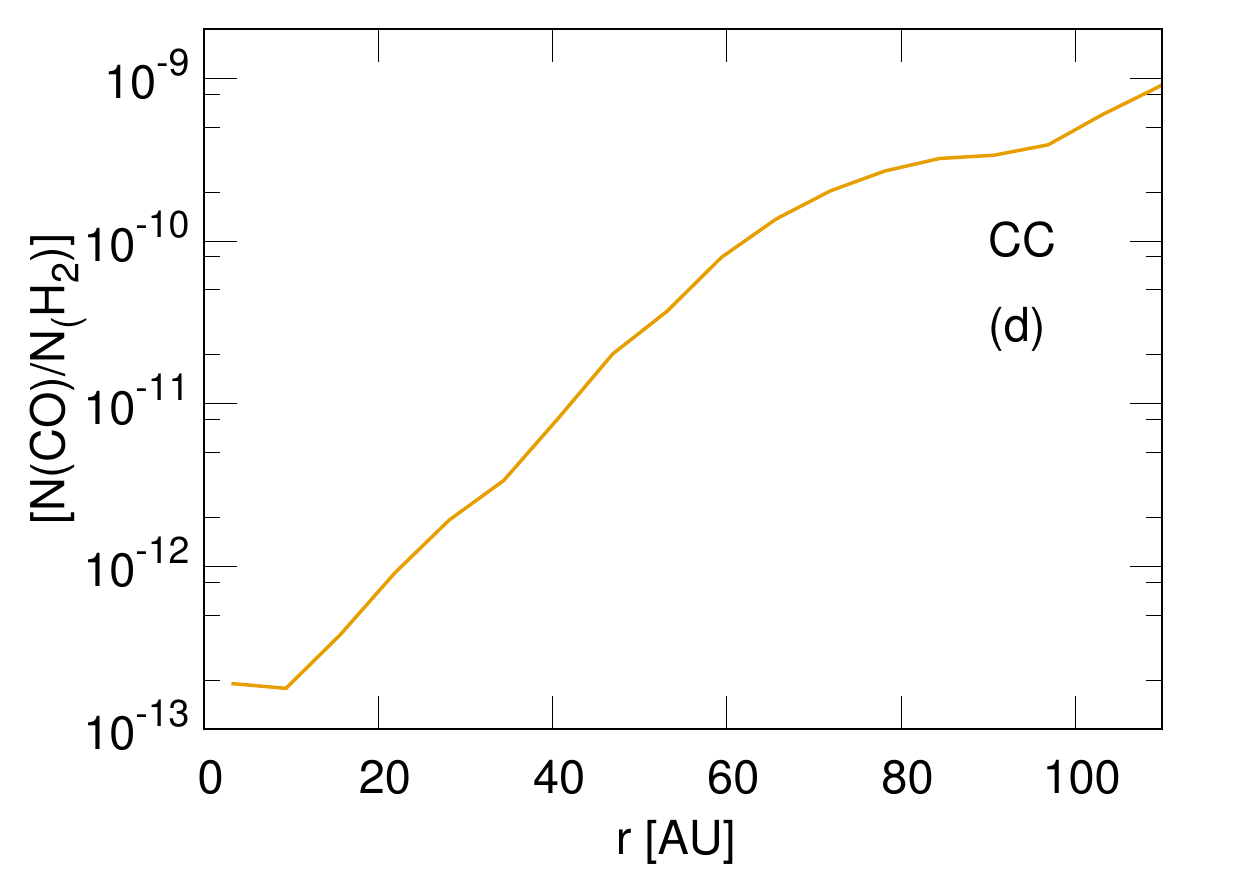}
     \caption{({\bf a,b,c}) The density, temperature, and velocity profiles at an evolutionary stage of 30,000-40,000 yr in the CC model. ({\bf d}) The CO abundance profile that provides the best-fit to the observed spectrum. The radial profiles are the azimuthal average along the line of sight.  }
     \label{CC-model-2}
  \end{figure*}

Without any artificial setting, four different components or zones appear naturally with the cloud evolution: (i) an infalling envelope that evolves into a circumstellar pseudo-disc; (ii) an inner Keplerian disc; (iii) a low-velocity outflow; (iv) a high-velocity jet. A schematic diagram of the model structure is shown in Fig.~\ref{CC-model-1}. The pseudo-disc is the inner regions of the infalling envelope and accretes onto the central proto-brown dwarf and the Keplerian disc that is embedded in it. It is a dynamically collapsing structure with little rotating motion (Sect.~\ref{intro}). The density, temperature, and velocity profiles for the CC model are shown in Fig.~\ref{CC-model-2}abc. The radial profiles are the azimuthal average along the line of sight. The sudden rise in the temperature for $r <$20 AU can be attributed to the presence of the high-velocity jet as well as some contribution from the stellar irradiation.

 \begin{figure*}
  \centering     
     \includegraphics[width=3.in]{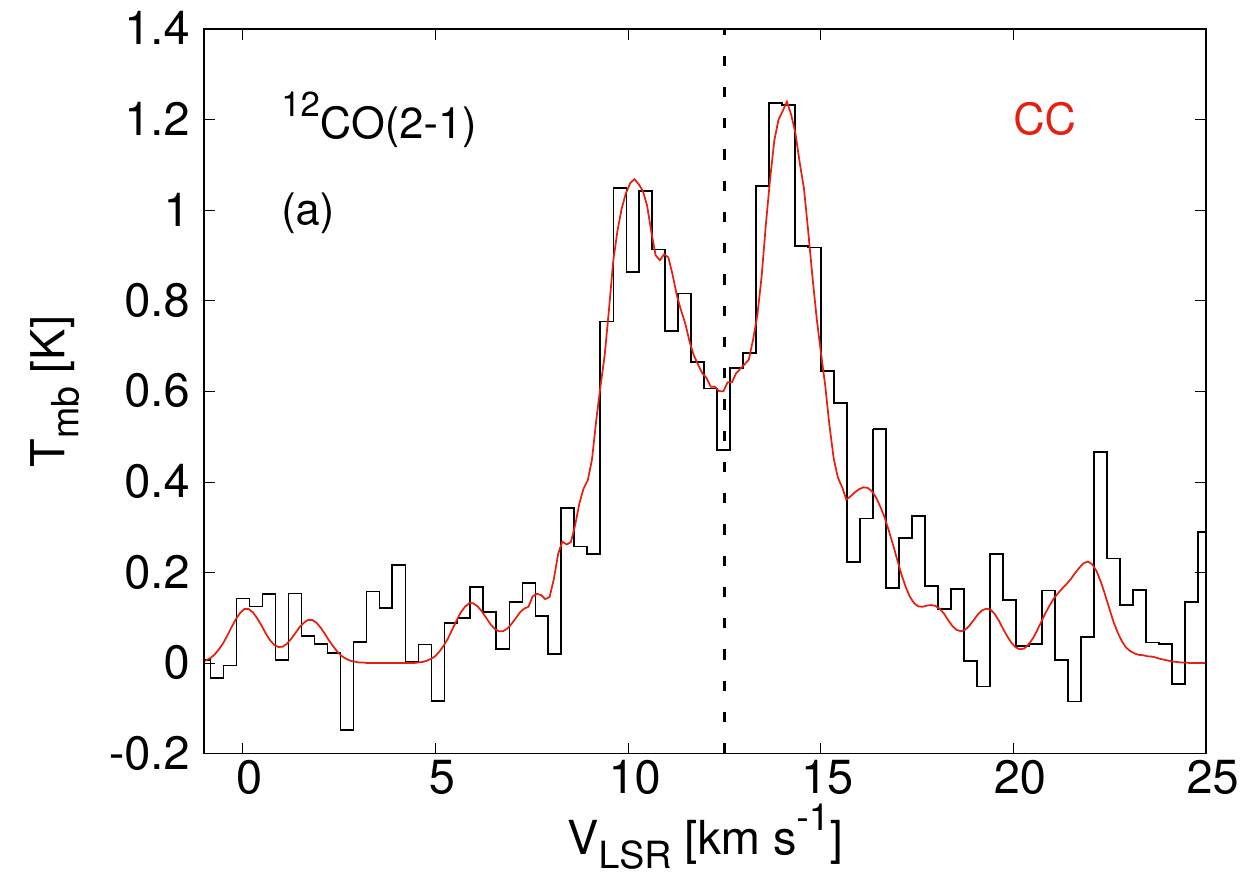}   \\
     \includegraphics[width=3.3in]{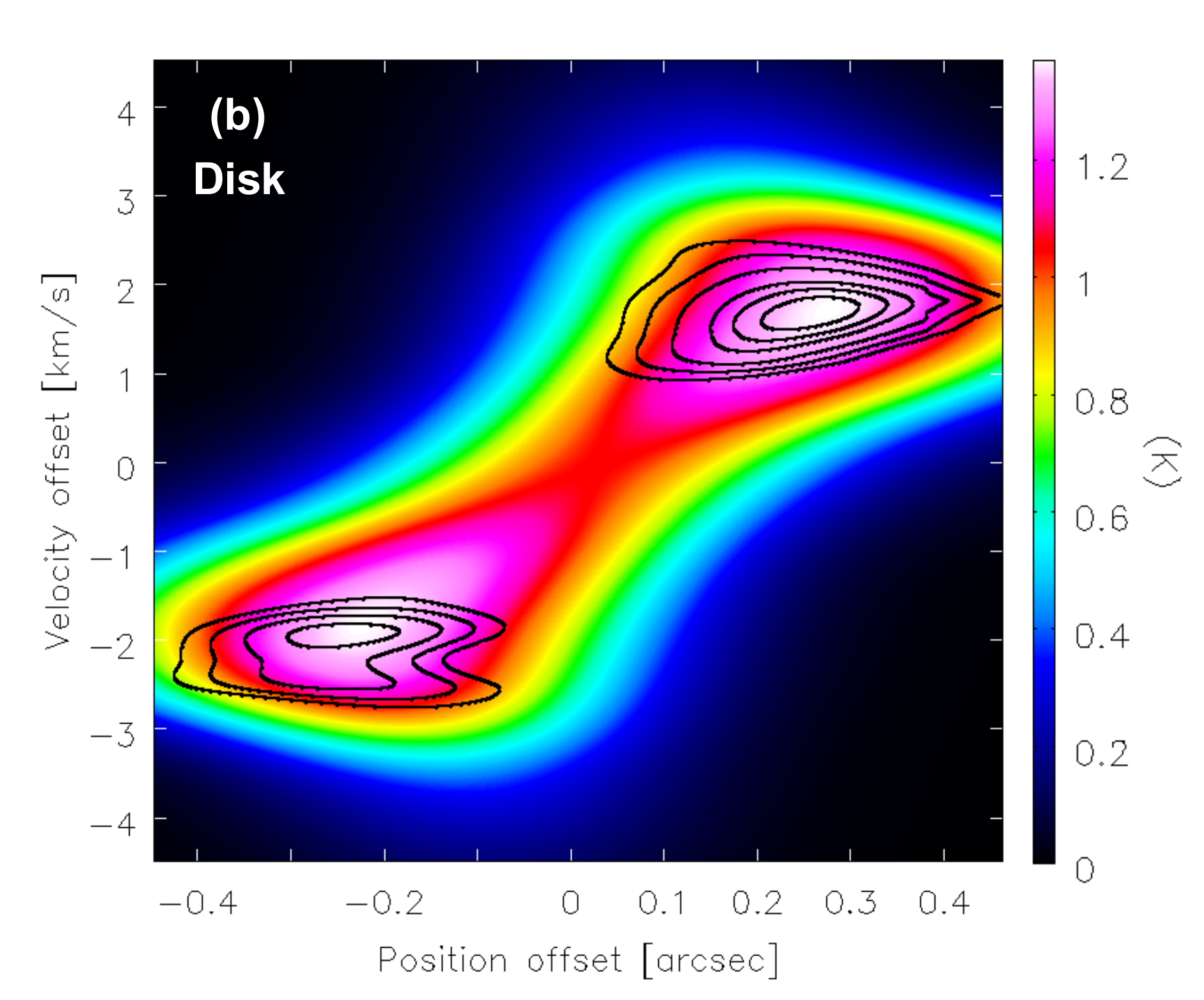}          
     \includegraphics[width=3.3in]{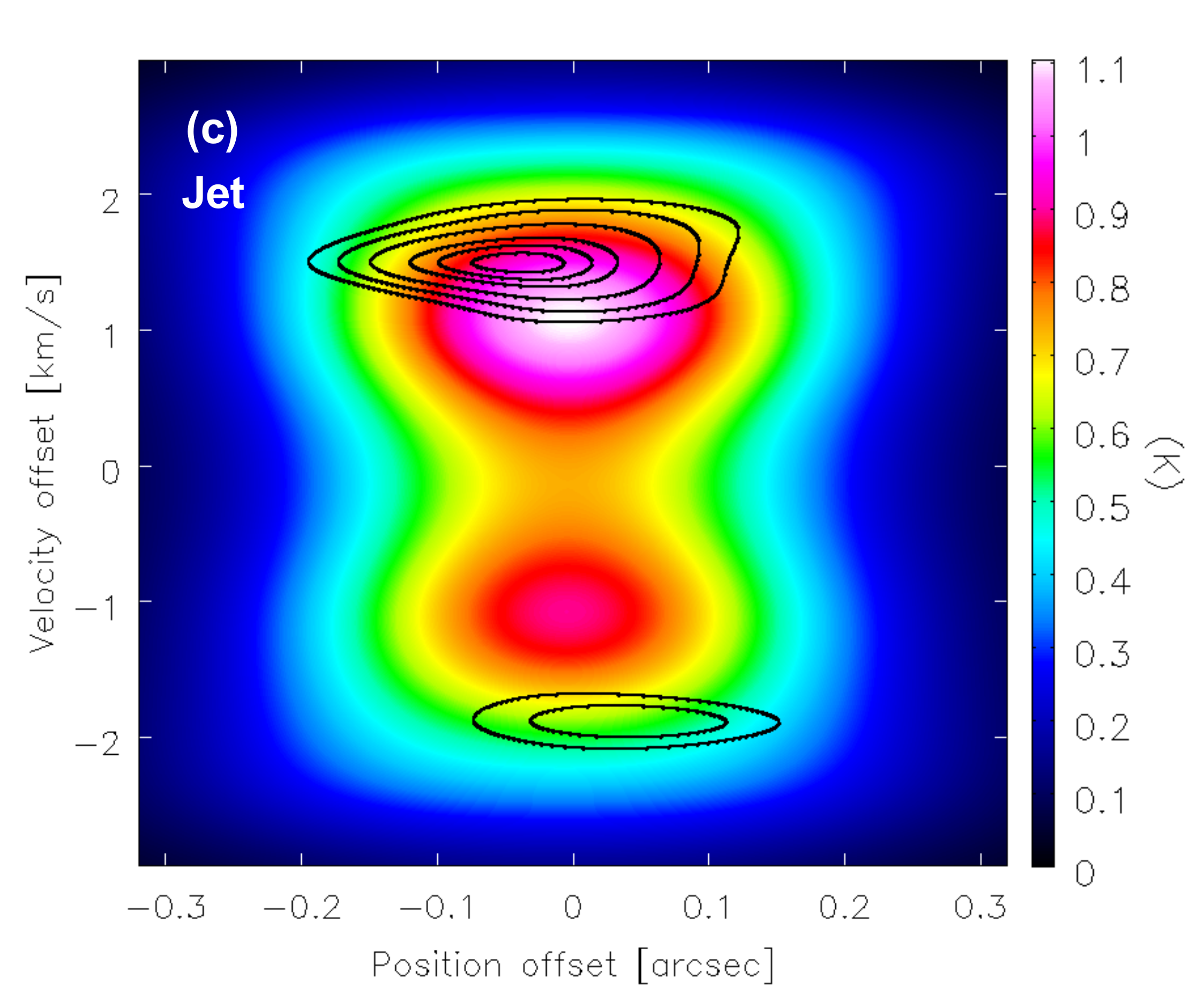}     
     \caption{({\bf a}) The CC model fit (red) to the observed CO spectrum (black). Dashed line marks the source V$_{LSR}$. ({\bf b}) Raster map shows the CC model PVD along the disc axis; colour bar shows the flux scale in units of K. Overplotted is the observed CO disc PVD in black contours. The contour levels are the same as in Fig.~\ref{pvd-co}. ({\bf c}) A comparison of the CC model PVD along the jet axis (raster map) with the observed CO jet PVD (black contours). }
     \label{CC-modelfit}
  \end{figure*}  

%In making the PVDs, emission brighter than 24 mJy is considered.  

%jetContours: 0.0273 0.0297 0.032 0.0344 0.0355 0.0365    
%diskContours: 0.0221 0.0241 0.0262 0.0282 0.0293 0.0301

Figure~\ref{CC-modelfit}a shows that the CC model can provide a good fit to both the asymmetric peaks and the broad wings in the CO spectrum. The best-fit was obtained for a close to edge-on inclination of 60$\degr$-70$\degr$. The reduced-$\chi^{2}$ value of the best-fit is 1.1. Figure~\ref{CC-model-1}d shows the radial profile of the CO abundance that provides the best-fit to the observed line profile. The abundance profile is basically the inverse of the density structure and shows severe depletion in the CO abundance towards the central densest regions of the proto-brown dwarf. The volume-averaged CO abundance relative to H$_{2}$, [N(CO)/N(H$_{2}$], obtained from the CC line fit is (5$\pm$2)$\times$10$^{-6}$.

%awk average is 1.68488e-09. 
%old : 8.6e-9
% Use the volume average abundance for 20 MJup disk mass, l=8 grid (282 au)-- this is comparable to the source size of 272 au :
%H2 Number Density= 6.879938e+08 cm^-3
%H2 Column Density= 1.422049e+24 cm^-2  --> NH2_gas
%Av_abund = 3.393125e-08

%From continuum (NH2_dust)  
% H2 number density = Mass (Msun) / (Volume * mu_H) = 0.02 * 1.898E30 / ( (4*PI/3) * (radius^3)*mu_H)
% H2 number density = 8.11E13 * 20 (MJup) / (135 AU ^ 3) = 6.92E8 cm^-3
% H2 Column density = nH * Renv = 6.92E8 * 135 au * 1.496E13 cm = 1.398E24 cm^-2

%Ndust/Ngas = 0.98. => 1-0.98 = 0.02 * 100 => only ~1.7\% of CO is depleted from the gas phase. 

Figure~\ref{CC-modelfit}b shows a comparison of the CC model and the observed disc PVD. The observed position and velocity peaks are well-matched with those predicted by the model. In particular, the model can re-produce the velocity spread of $\sim$2 km s$^{-1}$ seen in the observations. This velocity spread in the model is dependent on the evolutionary stage of the system. During the pre-stellar stage, the envelope is the dominant component and produces a small velocity spread of $<$0.5 km s$^{-1}$. As the system reaches the protostellar stage, there are, at least, three components of (rotation) disc, pseudo-disc, and infalling envelope. The infalling envelope and pseudo-disc are more massive than the Keplerian disc, and have different velocities, which creates a wide velocity spread of $\sim$2-4 km s$^{-1}$. As the brown dwarf system evolves, the infalling envelope and then the pseudo-disc gradually dissipate and their mass decreases, while the rotation disc grows in mass and size. %Thus, it is expected that the velocity spread becomes small again at later ($>$50,000 yr) stages of evolution. 

We checked for a match with the observed velocity offset using the CC simulation data at different evolutionary stages. From the best model fit to the PVD, we can constrain the age of the M1701117 system to be approximately 30,000-40,000 yr. The proto-brown dwarf is therefore not as young as the dynamical age of the HH~1165 jet ($\sim$2900 yr; Riaz et al. 2017), but has reached the protostellar phase ($\sim$0.01-0.1 Myr; e.g., Evans et al. 2009). As already noted in Sect.~\ref{source}, M1701117 is at an intermediate stage between Stage 0 and Stage I. As the CC simulations reach the $\sim$30,000-40,000 yr stage, the infalling envelope has flattened into the pseudo-disc structure (Fig.~\ref{CC-model-1}).

The predicted radius of the pseudo-disc from the CC model is about $\pm$50-100 AU. The total (dust+gas) mass encompassed within $\sim$200 AU considering the density in the range of 10$^{9}$ cm$^{-3}$ $< \rho <$ 10$^{12}$ cm$^{-3}$ is predicted to be $\sim$23 M$_{Jup}$, with a mass ratio $M_{circum}$/$M_{star} \sim$ 0.3. This circumstellar mass is dominated by the mass of the pseudo-disc. The physical dimension predicted by the CC model compares well with the spatial extent of about $\pm$77-96 AU for the pseudo-disc structure seen in the disc PVD (Fig.~\ref{pvd-co}), and the predicted circumstellar mass and $M_{circum}$/$M_{star}$ ratio are also consistent with that measured for the M1701117 system (Sect.~\ref{source}). The predicted size for the inner Keplerian disc is $<$5 AU, which is nearly impossible to observationally resolve at the distance to M1701117. Note that the predicted sizes are approximate as it is difficult to clearly identify the boundary between the components, particularly between the inner rotation disc and the pseudo-disc, since each physical component evolves differently with cloud evolution.

A comparison of the jet PVD with the CC model (Fig.~\ref{CC-modelfit}c) shows a similar asymmetry in brightness as seen in the observed case, and can be explained by the close to edge-on inclination of the system that could result in partial obstruction of the blue-shifted outflow lobe. There is a good match between the CC model and observations for the brighter red-shifted lobe, whereas the blue lobe is shifted by $\sim$1 km s$^{-1}$ and slightly towards the south-west than predicted by the model. The deflection in the blue lobe appears similar to that seen in the southern part of the HH~1165 jet close to the driving source (Riaz et al. 2017), and is possibly caused by the wind-outflow collision (Sect.~\ref{discussion}).

\subsubsection{Disc Fragmentation}
\label{fragment}

 \begin{figure*}
  \centering     
     \includegraphics[width=2.5in]{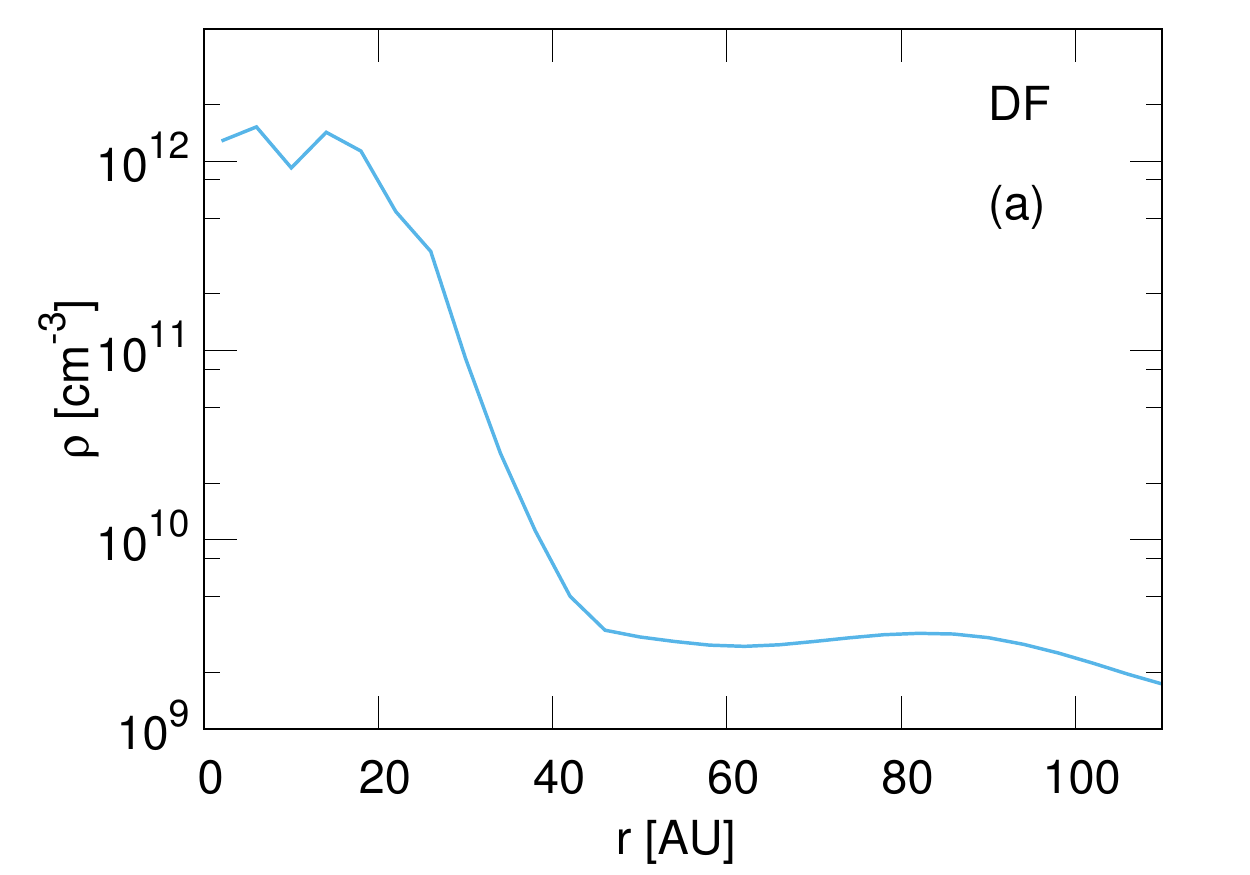}
     \includegraphics[width=2.5in]{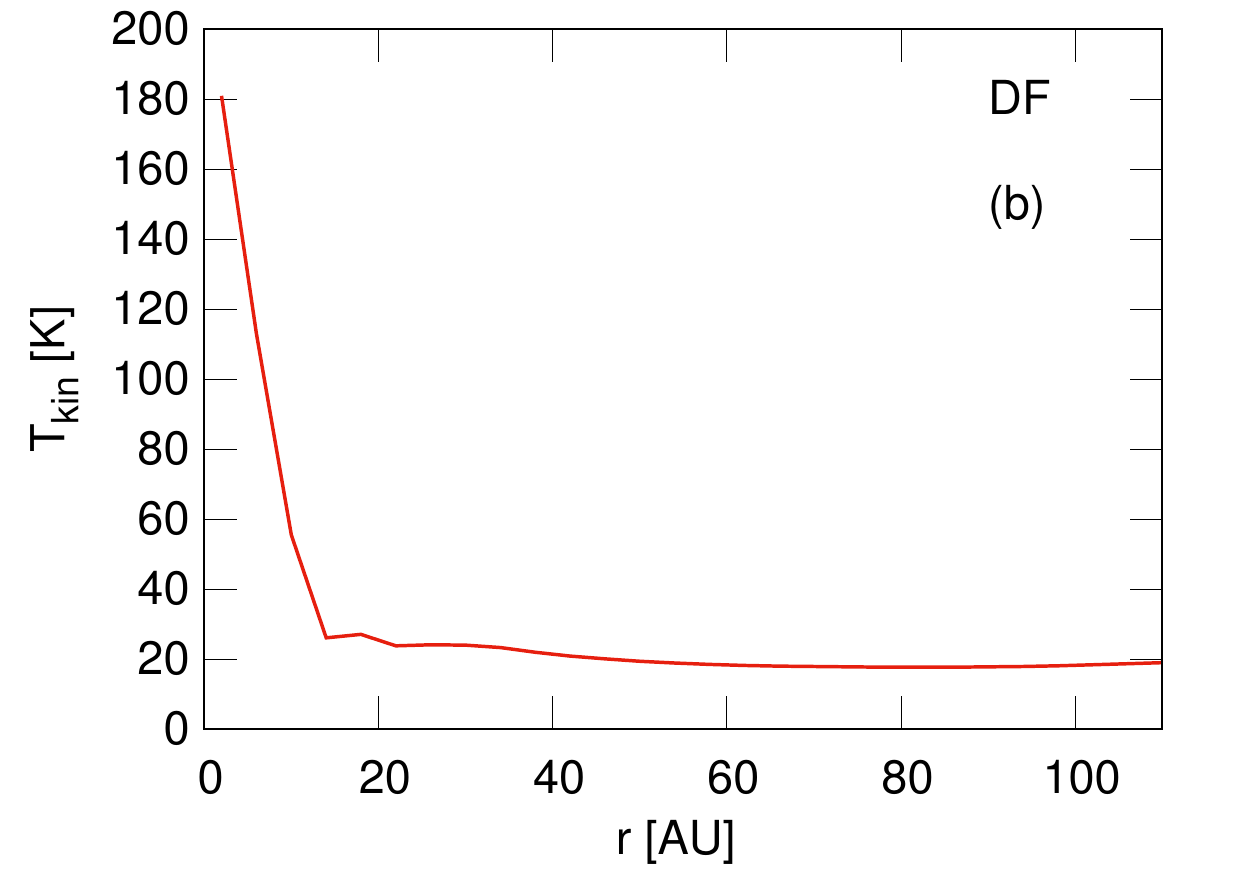}    \\
     \includegraphics[width=2.5in]{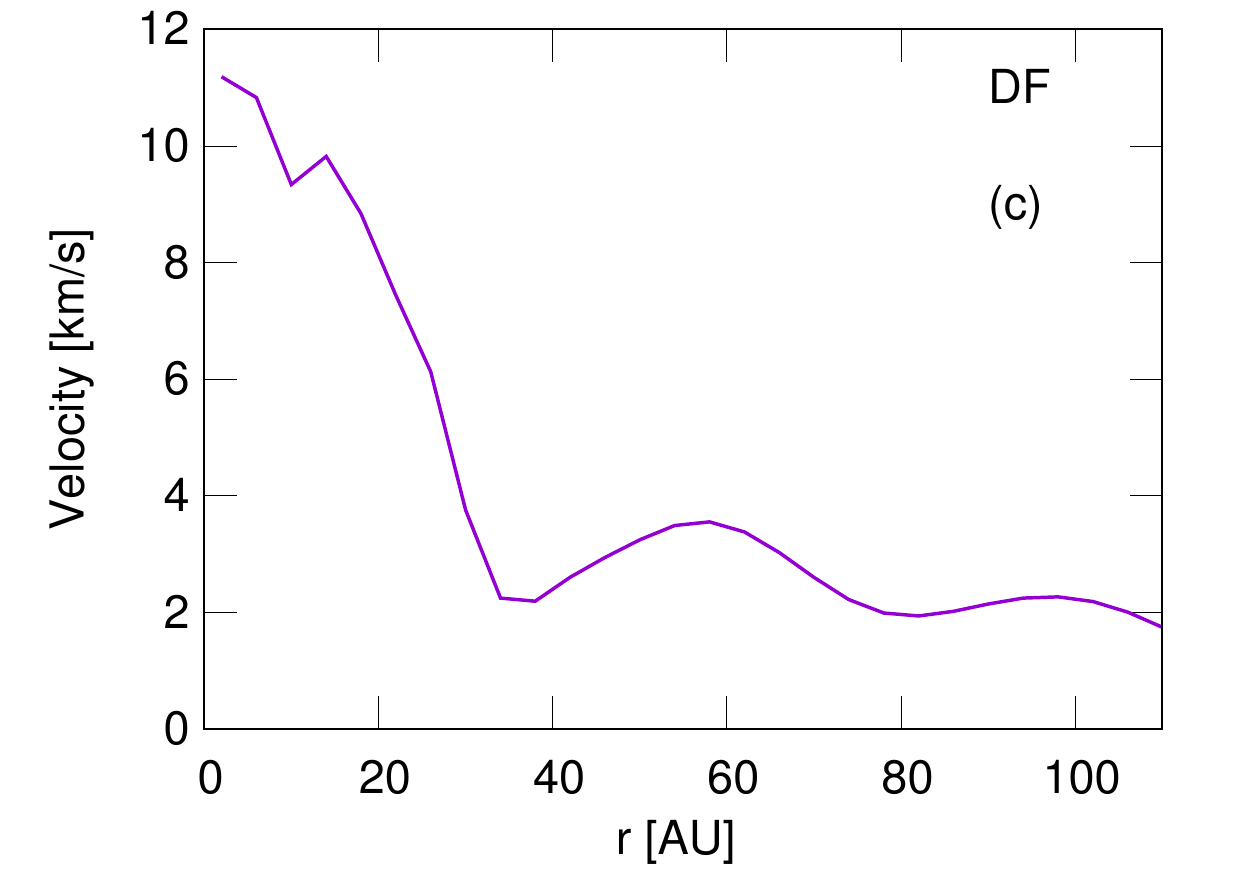}          
     \includegraphics[width=2.5in]{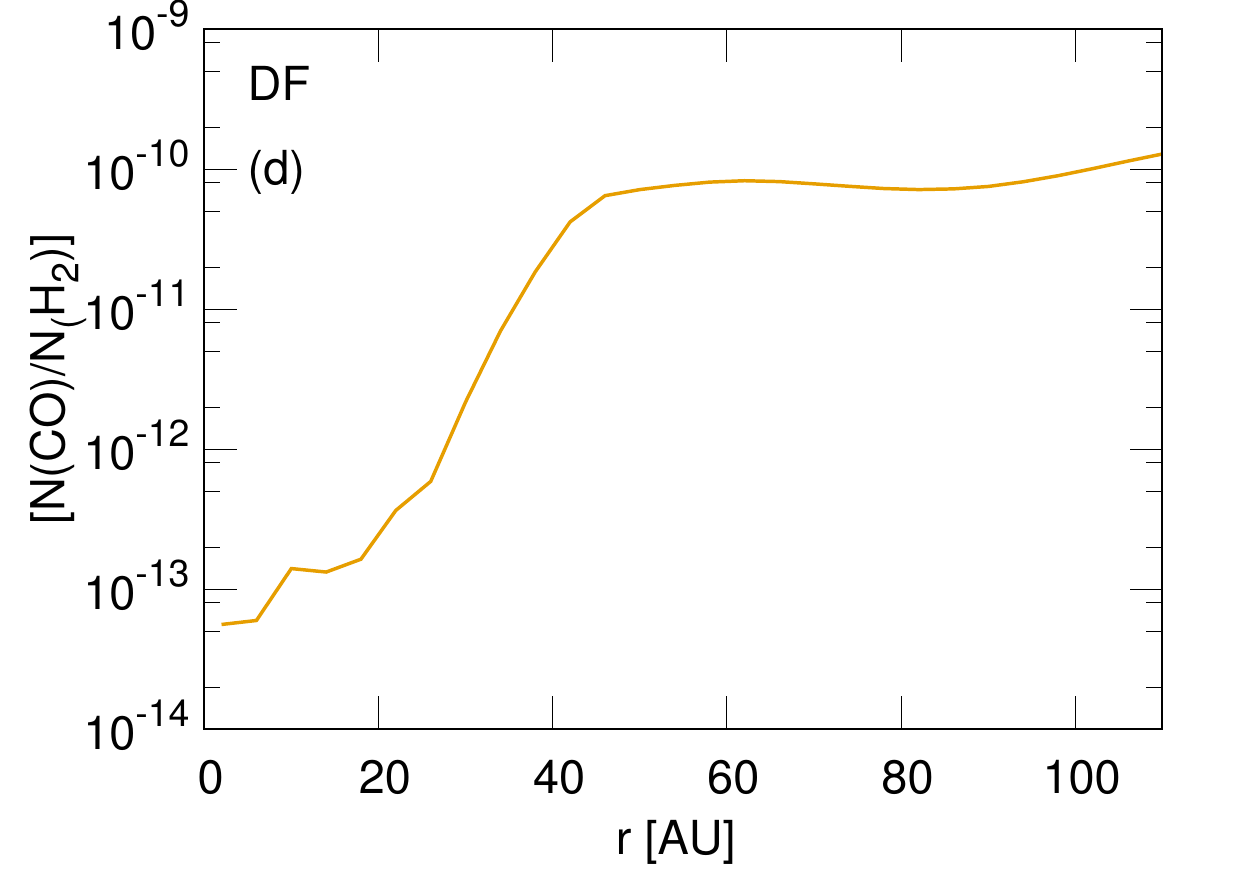}      
     \caption{({\bf a,b,c}) The density, temperature, and velocity profiles in the DF model. ({\bf d}) The CO abundance profile that provides the best-fit to the observed spectrum. The radial profiles are the azimuthal average along the line of sight.  }
     \label{DF-model}
  \end{figure*}

 \begin{figure*}
  \centering     
     \includegraphics[width=3.in]{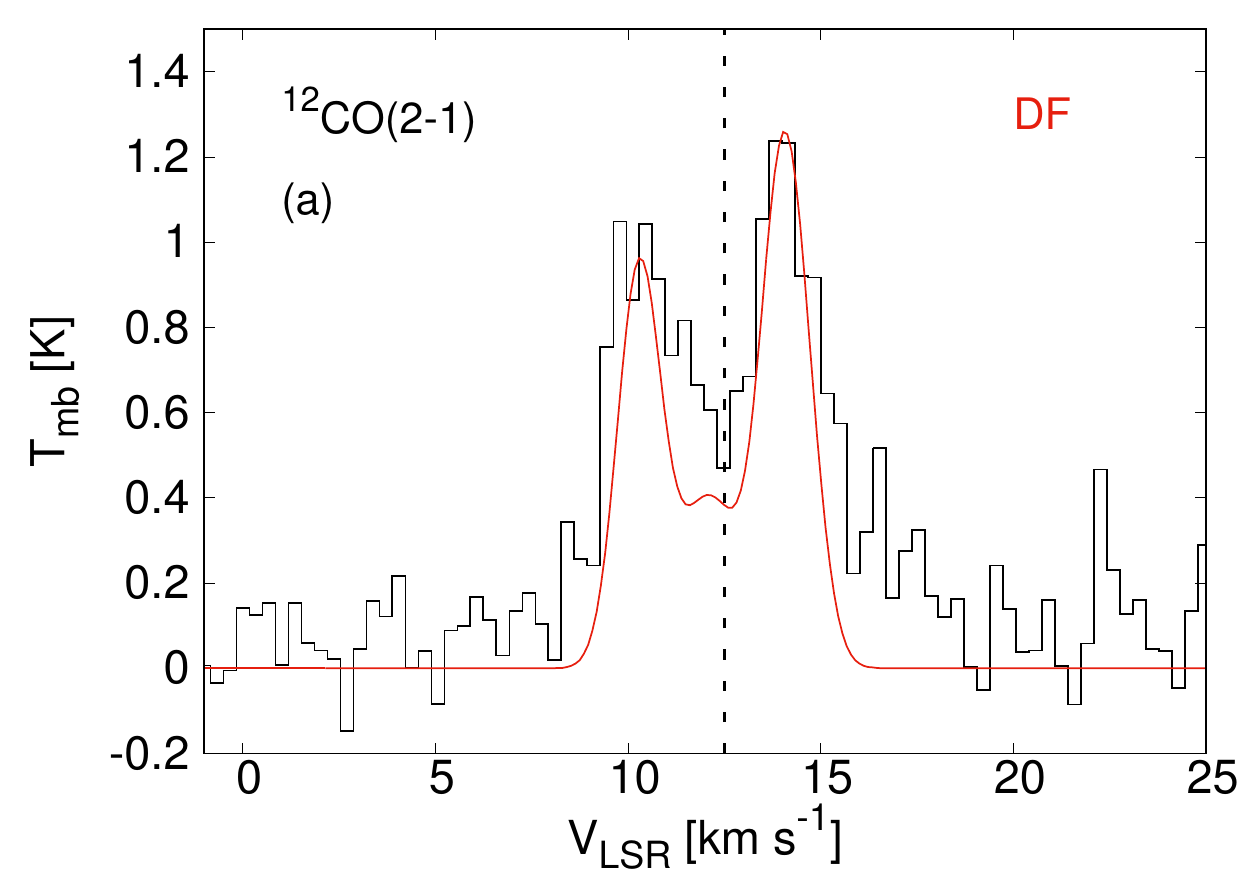}    
    \includegraphics[width=3.in]{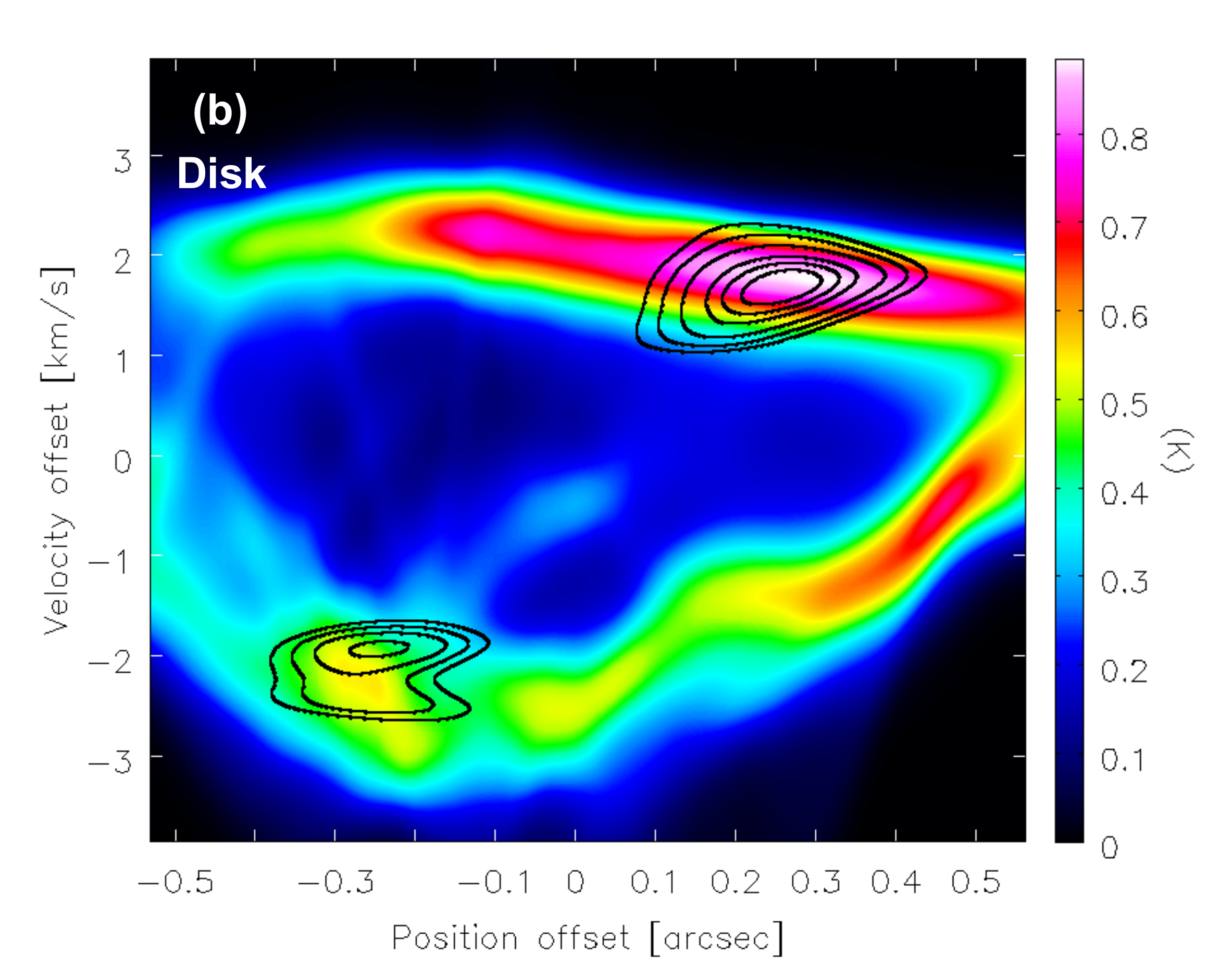} 
     \caption{({\bf a}) The DF model fit (red) to the observed CO spectrum (black). Dashed line marks the source V$_{LSR}$. ({\bf b}) Raster map shows the DF model PVD along the disc axis; colour bar shows the flux scale in units of K. Overplotted is the observed CO disc PVD in black contours. The contour levels are the same as in Fig.~\ref{pvd-co}. }
     \label{DF-modelfit}
  \end{figure*}  
  
%In making the PVDs, emission brighter than 24 mJy is considered.  
  
An alternative to a core collapse formation mechanism for brown dwarfs is considered to be the disc fragmentation model (hereafter, DF model), wherein the circumstellar disc around a massive star undergoes gravitational fragmentation resulting in the formation of very low-mass/sub-stellar cores that are then ejected from the system due to dynamical interactions (Stamatellos \& Whitworth 2009). We simulate a brown dwarf of 42~${\rm M_{Jup}}$ attended by a 20~${\rm M_{Jup}}$ disc. The simulations were performed using the Smoothed Particle Hydrodynamics code (SPH) {\sc Seren} (Hubber et al. 2011ab). The brown dwarf has formed by fragmentation in the disc of a solar-type star, followed by ejection. The details of the simulations are described in Stamatellos \& Whitworth (2009). The ejection in the DF simulations occurs at a stage of about 2-3 kyr and the evolution of the system is followed for another 2-3 kyr after ejection. We can therefore consider the ejected brown dwarf to be about the same age as the dynamical age of the HH~1165 jet.

The ejected brown dwarf possesses some circumstellar material in the form of a disc-like structure. This disc exhibits some asymmetries due to its violent dynamical history. This system has formed self-consistently within a larger-scale simulation of a fragmenting disc. The parameters of the system are the ones provided by the hydrodynamic simulation and have not been modified in any way to fit the observed object. It is difficult for a brown dwarf to sustain an infalling envelope when formed by disc fragmentation and subsequent ejection. The object may possess a thin envelope of a distorted shape, unlike a spherical or pseudo-disc structure seen for the CC model. Since outflows/jets are powered by the release of the gravitational energy of the accreting matter, it is expected that these components are weak for brown dwarfs formed via fragmentation/ejection. Note that the central brown dwarf will continue to accrete from the circumstellar disc but the accretion rate is expected to be quite weak, similar to the Class II brown dwarf accretion rate of the order of 10$^{-10}$ M$_{\sun}$/yr.

Figure~\ref{DF-model} shows the density, temperature, and velocity profiles for the DF model structure at the end of the simulations. The density profile (Fig.~\ref{DF-model}a) shows the high density material to be concentrated within $\sim$40 AU and then a sudden drop by $\sim$3 orders of magnitude at larger radii. Most brown dwarfs formed via fragmentation/ejection are expected to possess truncated discs with radii of $\leq$40 AU, although a few may possess larger sized discs (Stamatellos \& Whitworth 2009). When this brown dwarf+disc system is identified in the DF simulations, we selected a volume within $\sim$100 AU around it that contains diffuse gas from the larger-scale simulation ($\sim1~{\rm M_{Jup}}$) surrounding the ejected system. Most of this diffuse gas is concentrated near the disc midplane and exhibits a rotation at $\sim2\ {\rm km\ s^{-1}}$. Note that the density is still of the order of 10$^{9}$ cm$^{-3}$ for $r >$40 AU, making the $^{12}$CO (2-1) line with a critical density of 2$\times$10$^{4}$ cm$^{-3}$ easily detectable in the diffuse gas in outer disc regions.

The temperature profile (Fig.~\ref{DF-model}b) for an ejected brown dwarf shows comparatively warmer kinetic temperature of $\sim$20 K throughout the disc as compared to T$_{kin} \sim$7 K seen in the CC model. This is likely due to the multiple dynamical interactions that result in a warmer disc. The rise in T$_{kin}$ for $r <$15 AU is due to an increasing contribution from the stellar irradiation. The velocity profile for the DF model (Fig.~\ref{DF-model}c) shows wiggles at around 60 AU and 100 AU. The diffuse gas at $r >$40 AU is loosely bound to the system, which can result in such asymmetries in the velocity structure. The asymmetries are also expected due to the dynamics of the ejection that results in producing velocity differences in different regions of the disc (e.g. same radius but different azimuthal angle), as well as a large velocity spread of $\sim$2-4 km s$^{-1}$ in the outer regions of the disc.

%, due to which the optically thick region is within $\sim$40 AU and optically thin is in the outer regions

Figure~\ref{DF-modelfit}a shows the DF model fit to the observed CO spectrum. The spectral fit is a good match to the peaks but shows deeper self-absorption at the source V$_{LSR}$. The peaks are also narrower in the model than observed. Due to the absence of an outflow component, the extended wings in the observed spectrum cannot be reproduced by the model. The best-fit was obtained for an intermediate inclination of $\sim$50$\degr$. The reduced-$\chi^{2}$ value of the best-fit is 2.2. The abundance profile that provides the best-fit (Fig.~\ref{DF-model}d) clearly shows that any CO emission detected arises from the outer disc regions. The volume-averaged CO abundance relative to H$_{2}$, [N(CO)/N(H$_{2}$], obtained from the DF line fit is (1$\pm$0.5)$\times$10$^{-7}$, about an order of magnitude lower than that derived from the CC model fit. 

A comparison of the observed and DF modelled disc PVDs is shown in Fig.~\ref{DF-modelfit}b. The brighter, red-shifted lobe in the observations overlaps with the bright elongated feature seen in the model, while the blue lobe is much weaker in the model and appears as diffuse emission than a lobe. The wide spread in the model velocity profile (Fig.~\ref{DF-model}c) is a good match to the observed velocity shift. As noted, the DF model is expected to show asymmetries in the disc structure rather than a Keplerian-like profile. This is seen in the form of another bright elongated feature in the model at about -1 km s$^{-1}$ and +0.4$\arcsec$. The model PVD also shows a large halo devoid of gaseous material between $\sim$ -1 to 1 km s$^{-1}$ and $\pm$0.4$\arcsec$, which can explain the very low fluxes near the source V$_{LSR}$ in the model spectrum.

The DF model can re-produce the brightness asymmetry in the observations, but the shapes of the lobes are a poor fit. In Riaz et al. (2017), we had speculated based on the very young dynamical age of the HH~1165 jet that the M1701117 system may have been formed via the fragmentation of the disc around the nearby ($\sim$0.33 pc) massive B2IV-type star HR~1950, followed by ejection from the disc. While the asymmetries in the DF model cannot be modified or predicted, we have presented here just a single simulated case of a brown dwarf+disc formed via fragmentation/ejection. Modelling the observed PVD using a larger grid of DF simulations can provide a better understanding of how likely this formation scenario is for the M1701117 system.

\subsubsection{Single and Circum-binary Keplerian Discs}
\label{kepler}

 \begin{figure*}
  \centering       
     \includegraphics[width=2.7in]{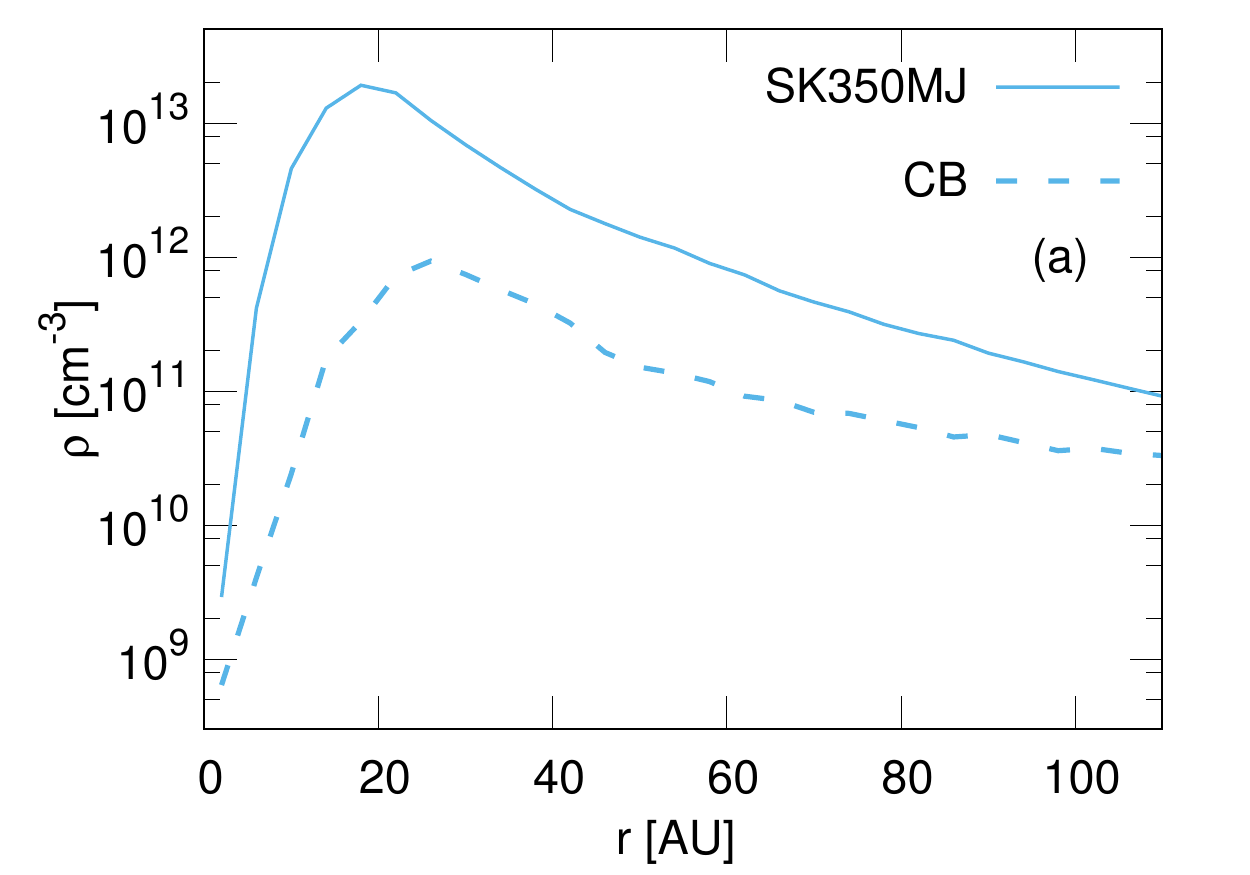}
     \includegraphics[width=2.7in]{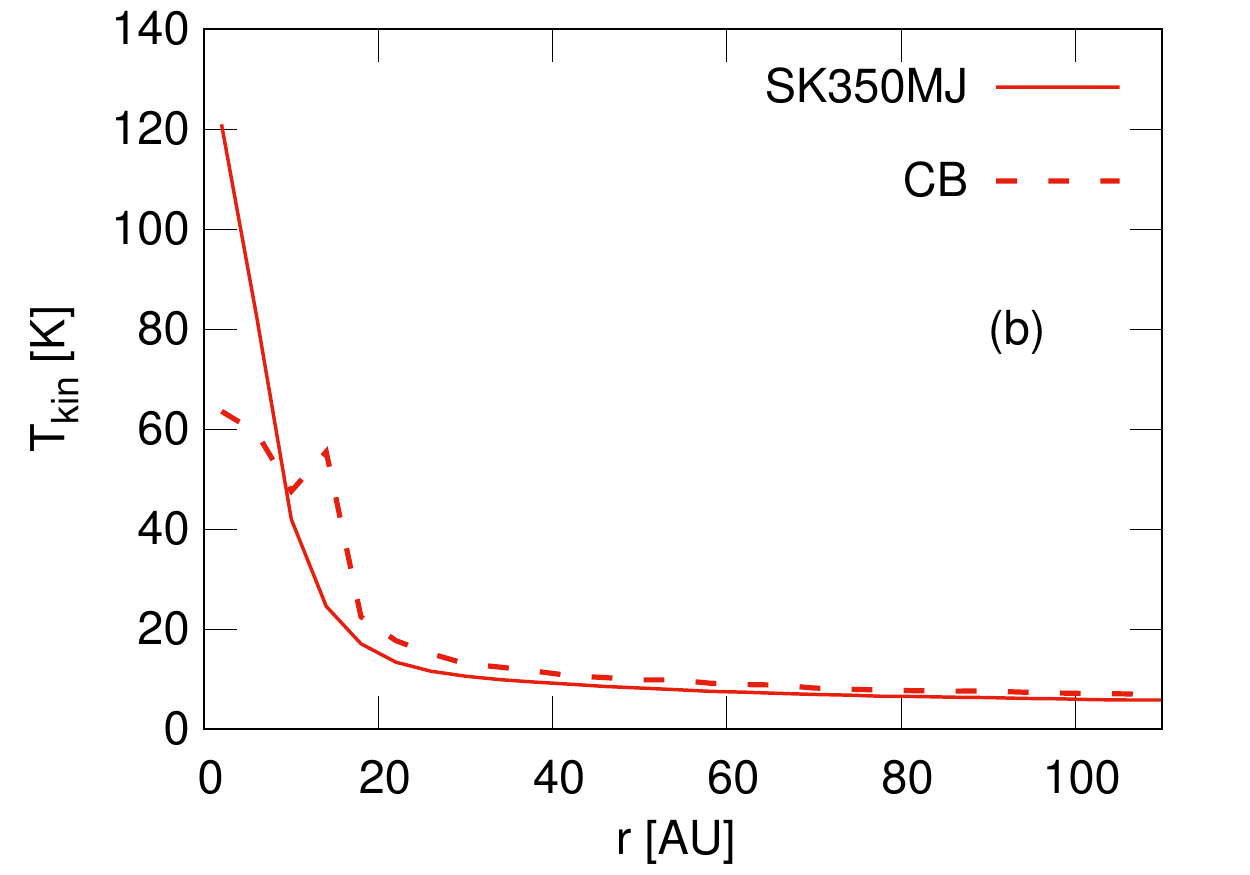}
     \includegraphics[width=2.7in]{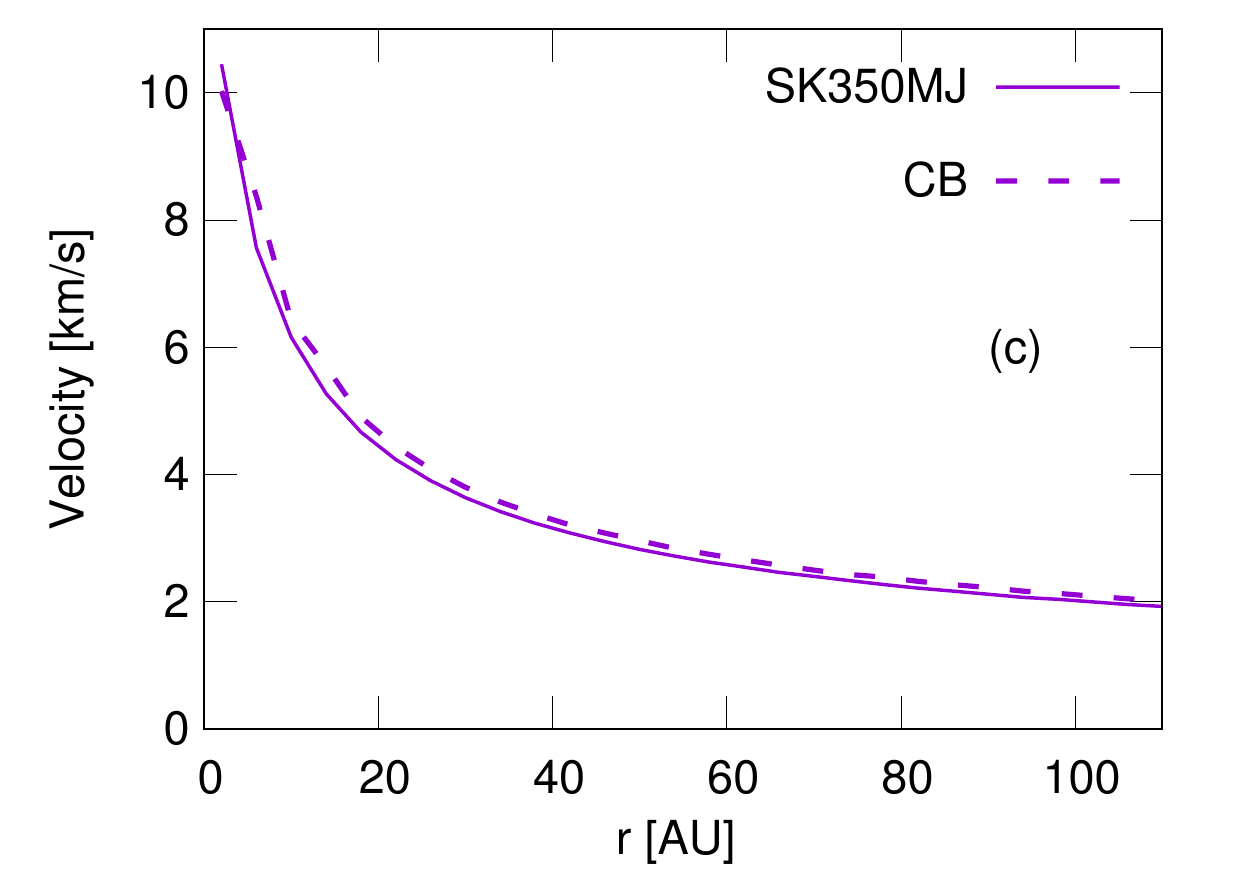} 
     \includegraphics[width=2.7in]{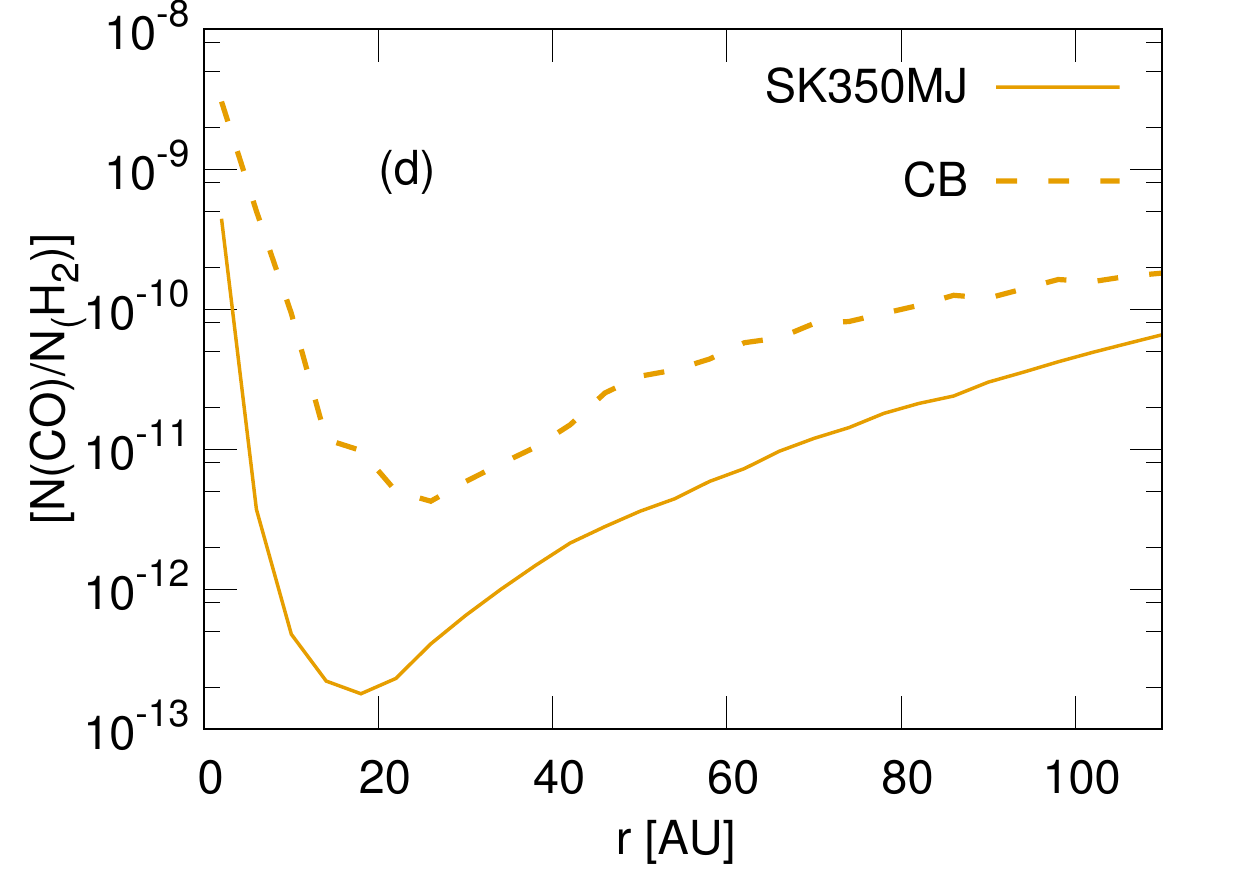}       
     \caption{({\bf a,b,c}) The density, temperature, and velocity profiles in the SK350MJ (solid line) and CB (dashed line) models. ({\bf d}) The CO abundance profiles that provide the best model fit to the observed spectrum. The radial profiles are the azimuthal average along the line of sight.  }
     \label{SK350-CB}
  \end{figure*}

 \begin{figure*}
  \centering       
     \includegraphics[width=3.2in]{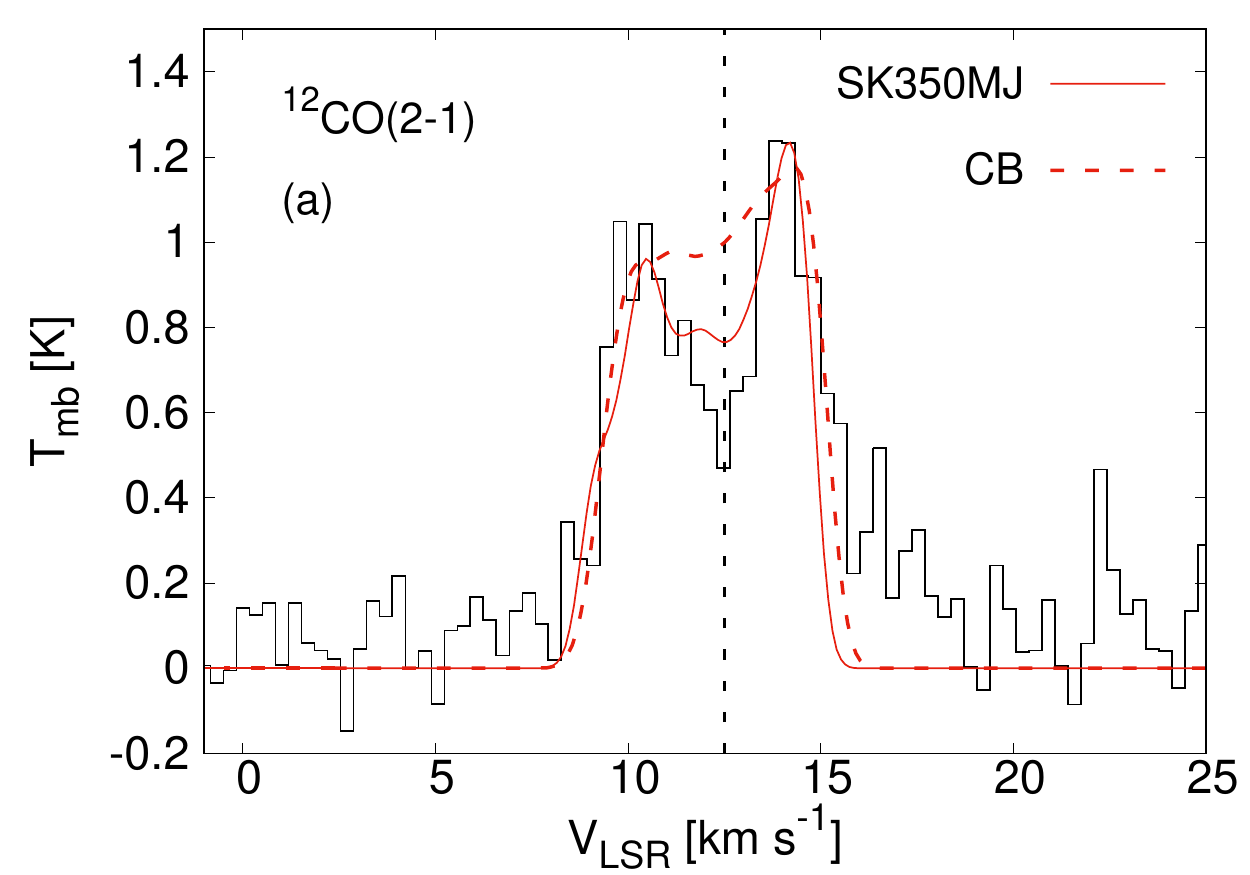}     
     \includegraphics[width=2.8in]{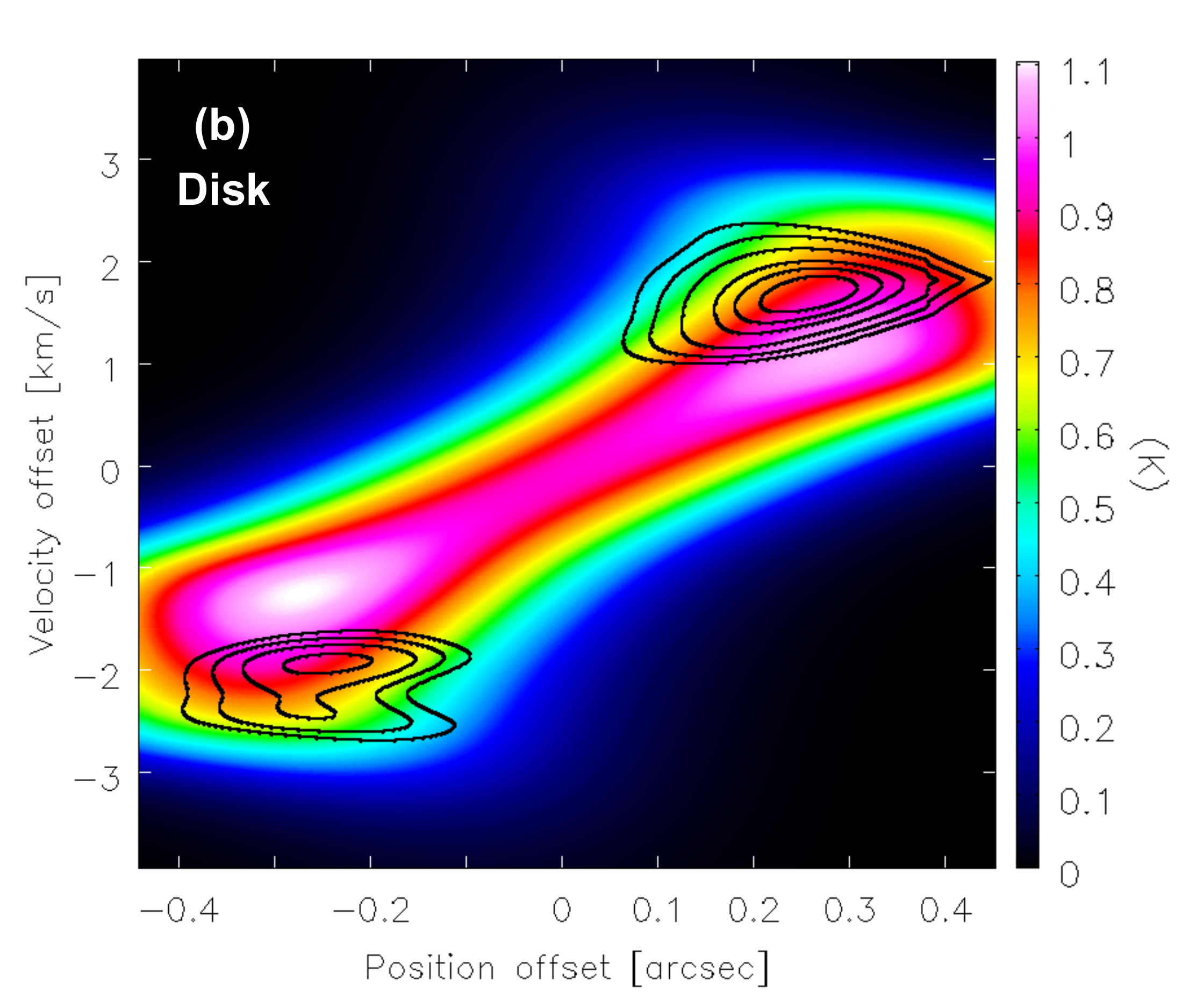}   
     \caption{({\bf a}) The SK350MJ (red) and CB (purple) model fits to the observed CO spectrum (black). Dashed line marks the source V$_{LSR}$. ({\bf b}) Raster map shows the SK350MJ and CB model PVD along the disc axis; colour bar shows the flux scale in units of K. Overplotted is the observed CO disc PVD in black contours. The contour levels are the same as in Fig.~\ref{pvd-co}. Both SK350MJ and CB models show the same structure in the disc PVD.  }
     \label{SK350-CB-fit}
  \end{figure*}       

%In making the PVDs, emission brighter than 24 mJy is considered. 

If we assume pure Keplerian kinematics, then the observed velocity shift of $\pm$2 km s$^{-1}$ and a $\sim$77-96 AU disc radius for M1701117 would imply a central object mass of $\sim$0.35 M$_{\sun}$, about 10 times higher than the stellar mass estimated for M1701117. We have simulated such a system of a 0.35 M$_{\sun}$ star surrounded by a 20~${\rm M_{Jup}}$ disc. It may also be the case that M1701117 is a binary system with a total central mass of 0.35 M$_{\sun}$ surrounded by a circum-binary disc of 20~${\rm M_{Jup}}$. In Riaz et al. (2017), we had noted a deflection in the southern lobe of the HH~1165 jet close to the driving source, such that the PA of the southern part is at an offset from the northern part of the jet. A similar offset is seen in the southern blue-shifted lobe of the CO outflow in the jet PVD (Fig.~\ref{pvd-co}). This hints towards a possible mis-alignment between the disc rotation axis and the outflow axis, which could produce an off-axis shock. Such mis-alignments have been seen in cases where the central driving source is a binary and the circumbinary disc is mis-aligned with the outflow (e.g., IRS43; Brinch et al. 2016). For M1701117, the near-infrared observations have a resolution of $\sim$0.1$\arcsec$ ($\sim$38 AU) and show no evidence of binarity. So this is likely a binary with a close ($<$38 AU) projected separation of the components. 

We have therefore tested these two scenarios for the case of M1701117 system. We simulated a disc around a primary object (single or binary) and let the disc evolve self-consistently to an equilibrium state. The simulations were performed using the SPH code {\sc Seren} (Hubber et al. 2011ab). The simulations take into account the disc self-gravity (which is computed using an octal-tree, as it is standard for SPH simulations), and the disc thermodynamics (using the method of Stamatellos et al. 2007). Therefore, the rotational velocity of the disc is nearly Keplerian, i.e. $v(R)=\sqrt{[G (M_\star+M_d(r<R))/R}$, where $M_d(r<R)$ is the disc mass within a radius $R$. The parameters we have used are: 

(i) a 350~${\rm M_{Jup}}$ M dwarf attended by a 20~${\rm M_{Jup}}$ disc, with an initial radius of 90 AU (hereafter, SK350MJ model), and

(ii) an M dwarf - M dwarf binary of total mass of 350~${\rm M_{Jup}}$, with equal mass components, attended by a 20~${\rm M_{Jup}}$ disc, with an initial radius of 90 AU (hereafter, CB model). The semi-major axis of the binary is set to 2 AU and the eccentricity to 0 (these may evolve as a result of the interaction between the binary and the disc).  

Figure~\ref{SK350-CB} shows the density, temperature, velocity profiles for these models, which are similar for the two cases. These discs expand with time and slowly accrete onto the central object. They are not embedded in a diffuse gas environment as the disc in the DF model. The central object(s) are represented by sink particles (Stamatellos et al. 2007), with radius of 1AU; gas that is within the sink radius and bound to the sink is then accreted. The use of sinks is needed to avoid small time steps in the hydrodynamic simulation but one drawback is that sinks tend to accrete gas at a higher rate than expected creating an gap near the central object. This is seen in Fig.~\ref{SK350-CB}a where the density starts to drop at $r<18$~AU. Therefore, the density near the central object may be underestimated in this case. In the run with the binary central object, the density starts to drop at $r<25$~AU. However, this is realistic as the binary is expected to carve out a central gap in the disc (Artymowicz et al. 1994; Stamatellos et al. 2018). The temperature closer to the central object(s) increases due to the heating by the stellar irradiation (Fig.~\ref{SK350-CB}b). The velocities reach $\geq$2 km s$^{-1}$ in the outer disc regions for both models (Fig.~\ref{SK350-CB}c).

Figure~\ref{SK350-CB-fit}a shows the model fits to the observed CO spectrum. None of the models can re-produce the self-absorption at the source V$_{LSR}$ seen in the spectrum. This is due to the high CO abundance produced by the models in the inner disc regions $r<$ 20 AU (Fig.~\ref{SK350-CB}d). The combination of a sudden rise in the kinetic temperature and a drop in the density for $r<$ 20 AU (Fig.~\ref{SK350-CB}ab) results in a rise in the CO abundance close to the central object(s), suggesting thermal desorption of the icy mantles and an increase the gas-phase CO abundance. An outflow component is not included in any of these models. Since the molecular outflow for M1701117 is unresolved, the effect is mainly seen in the extended wings of the spectrum and cannot be fit well by any of these models as compared to the CC model. The best-fits were obtained for an intermediate inclination of $\sim$50$\degr$-60$\degr$. The reduced-$\chi^{2}$ value of the best-fit is 2.3-2.5. A comparison of the observed disc PVD with these models (Fig.~\ref{SK350-CB-fit}b) shows an offset between the model and observations in both the red- and blue-shifted peaks, more notably for the blue one. The CO emission peaks in the models are at a smaller velocity offset at a given position than observed. The model PVD also shows a brightness gradient with a brighter blue lobe than the red-shifted one, which is possibly an inclination effect. Note that both SK350MJ and CB models show the same structure in the PVDs.

These models of isolated discs are meant to replicate the physical properties of the star+disc system in the present epoch. The poor model fits in Fig.~\ref{SK350-CB-fit} show that the structure of a 20 M$_{Jup}$ disc around a 0.35 M$_{\sun}$ single/binary star is inconsistent with the observed properties for the M1701117 system. The mismatch seen between the observed morphology and these models suggests that the input physical structure would require more components (envelope/outflow/jet) for a good match, in particular, the mismatch in the velocity structure in the PVDs (Fig.~\ref{SK350-CB-fit}b) is likely due to the absence of an infalling envelope. The density structure for these models is also notably different than the CC model. While it is difficult to determine which particular physical parameter causes the offset between the model and observed cases, the poor quality of these fits compared to the much refined CC model fit provide the justification that a purely Keplerian disc model cannot re-produce the observed morphology. Based on this, we can conclude that using Keplerian kinematics to infer the mass of the central object would be incorrect. Future high-resolution ($<$0.1$\arcsec$) observations can provide a better insight into the binarity scenario for M1701117.

\section{Discussion}
\label{discussion}

%{\bf We do not see a perfectly Keplerian morphology in terms of symmetric lobes with similar shapes, which are equidistant from the position and V$_{LSR}$ of the central source. This can be explained by the merged or unresolved emission from various components (infalling envelope, pseudo-disk, disk, jet, outflow) at the intermediate stage of the system that can produce differences in the morphology of the lobes.}

We have shown that the CC model that encompasses multiple components of an infalling envelope (in the shape of a pseudo-disc), an inner Keplerian disc, and a jet/outflow can provide a good fit to both the observed spectrum and the disc PVD for M1701117. The flattened envelope is infalling with rotation or spiralling onto the central proto-brown dwarf, and can be identified as a pseudo-disc in the CC simulations of brown dwarf formation. M1701117 is in transition from Stage 0 to Stage I, which is consistent with the finding of a pseudo-disc structure, intermediate between an infalling envelope and a Keplerian disc. A comparison of the volume averaged H$_{2}$ column density derived from CC modelling of the CO line (4.2$\times$10$^{22}$ cm$^{-2}$) with that derived from the 1.37 mm dust continuum emission (4.1$\times$10$^{22}$ cm$^{-2}$) indicates that only $\sim$2\% of the CO is depleted from the gas phase. This is unlike the high CO depletion fractions of $\sim$50-80\% seen in Class 0/I proto-brown dwarfs (Riaz et al. 2019), and could be due to the interaction of the jet with the inner dense envelope in the M1701117 system that can liberate CO molecules frozen onto the dust grains.

\begin{figure}
  \centering              
     \includegraphics[width=3in]{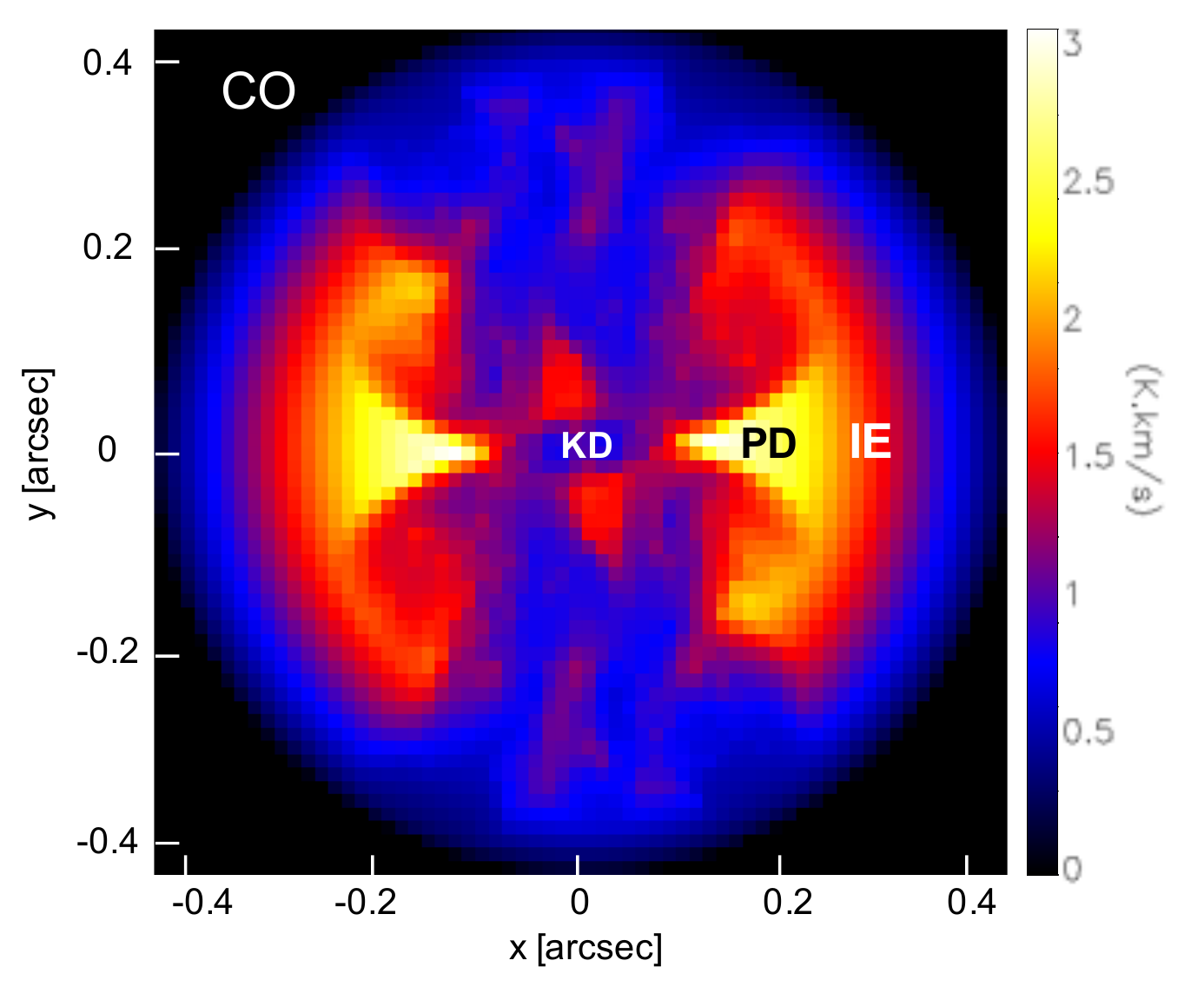}          
     \caption{The CC model predicted integrated intensity map in the CO (2-1) line derived from the best fit to the observed CO line profile. The map has not been convolved by the beam size. The labels `KD', `PD', and `IE' imply Keplerian disc, pseudo-disc, and the inner envelope regions, respectively. The colour bar shows the integrated flux scale in units of K.km s$^{-1}$.  }
     \label{model-map}
  \end{figure}

The bulk of the CO emission likely arises from the outer regions of the pseudo-disc, as suggested by the CC model integrated intensity map produced from the best fit to the observed CO spectrum (Fig.~\ref{model-map}). Note that this is the un-convolved map and thus shows details of physical components smaller than the angular resolution of the observations. The peak CO emission, as predicted by the model, arises from the outer edges of the pseudo-disc (labelled `PD'), with a contribution from the inner envelope (labelled `IE') region (Fig.~\ref{model-map}). The outer edge of the pseudo-disc, or the centrifugal radius, is where material from the collapsing envelope falls onto the disc. The region along the jet axis or perpendicular to the disc axis is nearly devoid of any molecular material and appears as a wide vertical cavity, except for a few shock emission knots or clumps formed due to the interaction of the jet with the inner envelope material. The model also predicts that CO is depleted only in the Keperian disc (labelled `KD'), which is the innermost ($<$0.05$\arcsec$) densest ($>$10$^{12}$ cm$^{-3}$) region embedded in the pseudo-disc (Fig.~\ref{model-map}). This is consistent with the fact that only about 2\% of CO appears to be frozen onto dust grains, as determined by comparing the dust and gas H$_{2}$ column densities.

Recently, there have been attempts made to estimate the gas mass independently from the dust mass by comparing the line luminosities of CO and its isotopologues with model grids (e.g., Ansdell et al. 2016; William \& Best 2014). For M1701117, the best model fit to the observed CO line comes from the core collapse model. In these simulations, the dust-to-gas ratio is fixed to 0.01, and it does not appear explicitly at different stages in the simulations. Therefore, it is difficult to determine if the ratio increases or decreases with the cloud evolution. Even though we have a good constraint now on the stage of the M1701117 system ($\sim$30,000-40,000 yr), we cannot reliably obtain an independent estimate on the gas mass from the model fit as that measurement implicitly assumes a gas to dust ratio of 100. We can speculate that if the dust size increases and the gas dissipates by outflow with time, then the dust-to-gas ratio would increase with time. Due to these uncertainties, we are unable to obtain an independent estimate on the gas mass derived from the CC model fit.

The deconvolved size of the pseudo-disc measured from the dust continuum image is 0.43$\pm$0.02$\arcsec$ or 165$\pm$7 AU. In comparison, the size measured from the CO line image is 0.5$\pm$0.05$\arcsec$ or 192$\pm$19 AU. The gas emission is therefore spread over a comparatively larger spatial scale than dust emission. Similar larger gas sizes compared to the dust sizes have been reported for discs around T Tauri and Herbig Ae/Be stars (e.g., Matthews et al. 2016; van der Marel 2016; Carney et al. 2018), and can be explained by the effects of photo-desorption due to UV radiation field (either internal or external) that can release the CO frozen onto dust grains into the gas phase, thus extending the line emission to larger spatial extents.

There is blue-shifted ($\sim$ -3 km s$^{-1}$) emission detected along the disc axis in H$_{2}$CO and N$_{2}$D$^{+}$ but at a different position offset. H$_{2}$CO peaks at the source location (Fig.~\ref{pvd-h2co}). M1701117 does not show emission in ortho-H$_{2}$CO but in para-H$_{2}$CO, suggesting an ortho-to-para ratio of $<$1. This is indicative of H$_{2}$CO formation on the surface of cold dust grains via ice surface reactions undergoing successive hydrogenation, and will result in enhanced H$_{2}$CO abundance where CO abundance is low (e.g., Kahane et al. 1984; Riaz et al. 2019). As shown through CC modelling, the CO abundance is lowest in the innermost, densest Keplerian disc region (Figs.~\ref{CC-model-2}d;~\ref{model-map}), suggesting that H$_{2}$CO emission likely probes the inner Keplerian disc where CO is expected to be frozen.

In comparison, N$_{2}$D$^{+}$ shows a slightly larger position offset (-0.36$\arcsec$) than seen for the blue-shifted lobe in the CO disc PVD, suggesting an origin from the outer regions in the pseudo-disc. N$_{2}$D$^{+}$ and H$_{2}$CO are both high-density tracers; the critical density of N$_{2}$D$^{+}$ (3-2) is 2.9$\times$10$^{6}$ cm$^{-3}$, and of para-H$_{2}$CO (3-2) is 9.7$\times$10$^{5}$ cm$^{-3}$. This is 1-2 orders of magnitude higher than CO (2$\times$10$^{4}$ cm$^{-3}$). The large position offset for N$_{2}$D$^{+}$ suggests that the emission in this line perhaps arises from a high-density clump being formed in the outer edges of the pseudo-disc. We adopted a spherical structure as the initial state in the CC model. However, initially distorted structure or turbulence may be possible to form N$_{2}$D$^{+}$ enhanced clump at the outer edge of the pseudo-disc. High-resolution imaging can provide some insight into the possible presence of a proto-planet in the making in the M1701117 pseudo-disc. It is also peculiar to see only blue-shifted emission in these lines, which is perhaps related to the CO brightness asymmetry; CO shows comparatively weaker emission in the blue lobe than the red one (Sect.~\ref{COlinePVDs}). N$_{2}$D$^{+}$ and H$_{2}$CO are known to peak in region where CO is relatively depleted in proto-brown dwarfs (Riaz et al. 2019a,b). Note that the H$_{2}$CO and N$_{2}$D$^{+}$ detection is quite weak ($\sim$2-3$\sigma$) in M1701117, therefore we can only speculate about the possible causes of the detection in these molecular lines.

We do not see symmetric strength and shape of the blue- and red-shifted lobes in the CO line PVD along the disc axis (Fig.~\ref{pvd-co}). We have explored the scenario if formation via fragmentation/ejection could produce the asymmetries in the pseudo-disc. Disks formed around ejected proto-brown dwarf embryos in the DF model tend to be asymmetric and distorted. This model can loosely re-produce the observed brightness asymmetry arising from the diffuse gas surrounding the proto-brown dwarf, but cannot provide a good fit to the shape of the lobes in the PVD. Note that the volume-averaged CO abundance derived from the DF model fit is an order of magnitude lower than that derived from the CC model fit, suggesting a possible difference in the molecular abundances between the two formation mechanisms. The CC model is a simulation devoted to the formation of a single brown dwarf whereas in the DF model the brown dwarf/disc system has formed due to gravitational instabilities in the disc around a solar-like star; therefore in the DF model a larger system is initially modelled, so that the resolution of the presented brown dwarf/disc system is not as high as in the CC model. Moreover the DF model does not include the effects of magnetic fields which lead to the formation of jets/outflows. Further DF models need to be developed that can simulate the range in disc asymmetries and re-produce high velocity shifts in disks around brown dwarfs. Also needed are models that include the outflow contribution and how that affects the fragmentation/ejected scenario. This can provide a better understanding of the differences in the chemistry of the brown dwarfs formed via these mechanisms.

%The stage of the system is also expected to be much younger than the tens of thousands of years estimated from the CC model fit. 

%{\bf The effects of photo-evaporation are already seen in the southern lobe of the HH~1165 jet that is in close proximity to a massive B-type star (Riaz et al. 2017). }

The asymmetry in brightness and morphology between the blue- and red-shifted lobes observed in the CO disc PVD (Fig.~\ref{pvd-co}) could be due to external irradiation effects or wind-outflow collision. M1701117 is located in an irradiated environment, the main source of which is the B2 type star HR~1950 that lies $\sim$0.33 pc away. The HH~1165 jet shows a curved structure aligned in a way that places the southern part of the jet closer to HR~1950 while the northern part is diverted away from it (Riaz et al. 2017). The Lyman continuum and He I continuum photon emission rate from a B2 star is estimated to be approximately 10$^{47}$ photons/s (e.g., Sternberg et al. 2003). The weaker outflow emission in the blue-shifted, south-east lobe in the CO jet PVD (Fig.~\ref{pvd-co}) is consistent with the faint detection in the optical [SII] $\lambda$6731 \AA ~emission seen towards the south-east part of the HH~1165 jet (Riaz et al. 2017), and can be attributed to the higher extent of photo-evaporation effect due to the proximity with BR~1950. We also see an offset along the jet axis between the position, velocity of the southern, blue-shifted lobe and the CC model (Fig.~\ref{CC-modelfit}b). A similar offset was seen close to the driving source in the southern part of the HH~1165 jet (Riaz et al. 2017), which could also be due to a deflection caused by the wind-outflow collision. Stellar winds from HR 1950 collide with the outflow and produce the observed jet/outflow deflection.

The best observable diagnostic of the external effects of irradiation from a nearby massive star is the circumstellar mass. In a sub-millimeter (856 $\mu$m) continuum survey of T Tauri stars in the Orion Nebula Cluster, a general trend is seen of a decline in the disc masses with decreasing projected distance from the massive star $\theta^{1}$ Ori C (Mann et al. 2014). The closer the T Tauri star to $\theta^{1}$ Ori C, the lower the disc mass due to the stronger effect of photo-evaporation. The trend is more notable for objects within $\sim$0.05 pc of $\theta^{1}$ Ori C that have disc masses of $\leq$3 M$_{Jup}$, while objects that are $>$0.5 pc from the massive star have disc masses of $\sim$10-100 M$_{Jup}$. For the case of M1701117 that lies $\sim$0.33 pc from HR~1950, the effects of photo-evaporation that may have significantly dissipated the circumstellar mass appear negligible. Nevertheless, the fact that molecular line emission in only detected at or very close ($<$100 AU) to the M1701117 source position, and none of the HH~1165 shock emission knots show any molecular line emission indicates that this region is mostly devoid of gaseous material.

\section{Summary}

We have analyzed the ALMA $^{12}$CO (2-1) line and 1.37 mm continuum observations at an angular resolution of $\sim$0.4$\arcsec$ for the proto-brown dwarf Mayrit 1701117, the driving source of the large-scale HH~1165 jet. The morphology seen in the CO disc PVD is suggestive of an origin from a pseudo-disc structure around the central proto-brown dwarf. It is peculiar to see a high velocity shift of $\pm$2 km s$^{-1}$, which for a pure Keplerian disc of $\sim$100 AU radius would imply a central object mass that is $\sim$10 times higher than the stellar mass estimated for M1701117. However, using a disc-only model for a 0.35 M$_{\sun}$ single or binary object(s) where the rotational velocity of the disc is pure Keplerian is a poor fit to both the observed CO spectrum and the disc PVD for M1701117. It would therefore be incorrect to infer on the mass of the central object by assuming a disc-only structure. On the other hand, the CC simulations can explain the observed large velocity spread of $\pm$2 km s$^{-1}$ in the outer pseudo-disc regions due to the combined emission from various components (infalling envelope, pseudo-disc, inner Keplerian disc, outflow) of the system. Future high angular resolution observations with ALMA can help distinguish between the inner Keplerian disc and the outer pseudo-disc regions, and provide a better insight into the contribution from the individual components.

%The source size for M1701117 as measured from the 1.37 mm continuum image is $\sim$270 AU. 
%The CO line PVD along the jet axis shows a compact ($<$150 AU) unresolved molecular outflow. 

%The very small value of the dust opacity index $\beta$$\sim$0.1 measured for M1701117 can be explained by the effect of dust grain growth.

%This object is in transition from Class 0 to Class I, consistent with the finding of a pseudo-disk structure, intermediate between an infalling envelope and a Keplerian disk. 

%the observed CO spectrum shows both infall and Keplerian rotation signatures as well as broad, extended wings indicative of an unresolved molecular outflow. 

%the observed continuum and line emission likely arises from a pseudo-disk component, intermediate between an infalling envelope and a Keplerian disk. 

%We have derived a CO abundance (relative to H$_{2}$) of 8.6$\times$10$^{-9}$ from line radiative transfer modelling of the observed CO spectrum. A comparison of the H$_{2}$ column density derived from the CO gas emission with that derived from 1.37 mm dust continuum emission indicates that $\sim$15\% of the CO is depleted from the gas phase. 

\section*{Acknowledgements}

This paper makes use of the following ALMA data: ADS/JAO.ALMA\#2016.1.01453.S. ALMA is a partnership of ESO (representing its member states), NSF (USA) and NINS (Japan), together with NRC (Canada), MOST and ASIAA (Taiwan), and KASI (Republic of Korea), in cooperation with the Republic of Chile. The Joint ALMA Observatory is operated by ESO, AUI/NRAO and NAOJ. BR would like to thank Thomas Stanke, Tom Booth, and Sarah Wood for help with the data reduction, Susan Terebey and Catherine Dougados for their comments and suggestions, and Nicolas Lodieu for providing the UKIDSS data in $\sigma$ Orionis. BR acknowledges funding from the Deutsche Forschungsgemeinschaft (DFG) - Projekt number 402837297.

%%%%%%%%%%%%%%%%%%%%%%%%%%%%%%%%%%%%%%%%%%%%%%%%%%

%%%%%%%%%%%%%%%%%%%% REFERENCES %%%%%%%%%%%%%%%%%%

%%%%%%%%%%%%%%%%% APPENDICES %%%%%%%%%%%%%%%%%%%%%

%\appendix

%\section{Some extra material

%%%%%%%%%%%%%%%%%%%%%%%%%%%%%%%%%%%%%%%%%%%%%%%%%%

% Don't change these lines
\bsp	% typesetting comment
\label{lastpage}
\end{document}